\documentclass[journal]{new-aiaa}
\usepackage[utf8]{inputenc}
\DeclareUnicodeCharacter{2212}{-}
\usepackage{textcomp}

\usepackage{graphicx}
\usepackage{subfig}
\usepackage{cleveref}
\usepackage{enumitem}
\usepackage{multicol,multirow}
\usepackage{amsmath,amsfonts}
\usepackage{siunitx}
\usepackage{longtable,tabularx}
\usepackage{soul}
\usepackage{ulem}
\usepackage{algorithm}
\usepackage{algorithmic}

\title{Data--Driven Turbulence Modeling Approach for Cold--Wall Hypersonic Boundary Layers}

\author{Muhammad I. Zafar \footnote{Ph.D. Student, Aerospace and Ocean Engineering. Member AIAA; zmuhammadirfan@vt.edu (Corresponding Author)}, 
Xuhui Zhou\footnote{Ph.D. Student, Aerospace and Ocean Engineering.}, 
and Christopher J. Roy\footnote{Professor, Aerospace and Ocean Engineering. Associate Fellow AIAA.}}
\affil{Virginia Polytechnic Institute and State University, Blacksburg, Virginia 24061}

\author{David Stelter\footnote{Group Leader, Machine Learning and Data Analytics.}}
\affil{Spectral Sciences Inc., Burlington, Massachusetts 01803}
\author{Heng Xiao\footnote{Professor, Stuttgart Center for Simulation Science (SimTech).}}
\affil{University of Stuttgart, Stuttgart, Germany}

\begin{document}

\maketitle

\begin{abstract}
Wall--cooling effect in hypersonic boundary layers can significantly alter the near--wall turbulence behavior, which is not accurately modeled by traditional RANS turbulence models. To address this shortcoming, this paper presents a turbulence modeling approach for hypersonic flows with cold--wall conditions using an iterative ensemble Kalman method. Specifically, a neural--network--based turbulence model is used to provide closure mapping from mean flow quantities to Reynolds stress as well as a variable turbulent Prandtl number. Sparse observation data of velocity and temperature are used to train the turbulence model. This approach is analyzed using direct numerical simulation database for  zero-pressure gradient (ZPG) boundary layer flows over a flat plate with a Mach number between 6 and 14 and wall--to--recovery temperature ratios ranging from 0.18 to 0.76. Two training cases are conducted: 1) a single training case with observation data from one flow case, 2) a joint training case where data from two flow cases are simultaneously used for training. Trained models are also tested for generalizability on the remaining flow cases in each of the training cases.
The results are also analyzed for insights to inform the future work towards enhancing the generalizability of the learned turbulence model.
\end{abstract}

\section*{Nomenclature}

{\renewcommand\arraystretch{1.0}
\noindent\begin{longtable*}{@{}l @{\quad=\quad} l@{}}
$C_f$ & $\tau_w /\left(\frac{1}{2} \rho_{\infty} U_{\infty}^2\right)$; wall skin friction coefficient \\
$C_{h,aw}$ & $q_w/(\rho_\infty c_p U_\infty (T_r - T_w))$; wall heat transfer coefficient \\
$C_{h,e}$ & $q_w/(\rho_\infty c_p U_\infty (T_{0,e} - T_w))$; wall heat transfer coefficient \\
$c_p$ & heat capacity at constant pressure, J/(K.kg) \\
$E$ & total energy, J \\
$f$ & generic solution variable \\
$F_s$ & factor of safety \\
$g^{(n)}$ & scalar coefficient of $n$--th tensor basis \\
$H$ & total enthalpy, J \\
$J$ & cost function \\
$K$ & Kalman gain matrix \\
$k$ & turbulent kinetic energy per unit mass, $k \equiv u_i^{\prime \prime} u_i^{\prime \prime} / 2$, J/kg \\
$k_T$ & thermal conductivity, W/(m$\cdot$K) \\
$M$ & Mach number \\
$\mathrm{P}$ & model error covariance matrix \\
$p$ & pressure, Pa \\
$Pr$ & 0.71; molecular Prandtl number \\
$Pr_t$ & turbulent Prandtl number \\
$q_j$ & local heat flux component in $x_j$ direction , W/m$^2$ \\
$q_w$ & wall heat flux, W/m$^2$ \\
$\mathrm{R}$ & observation error covariance matrix \\
$Re_\tau$ & $\rho_w u_\tau \delta / \mu_w$; Reynolds number based on friction velocity and wall viscosity  \\
$\mathbf{S}$ & $\frac{1}{2} \left( \frac{\partial u_i}{\partial x_j} + \frac{\partial u_j}{\partial x_i} \right)$; strain rate tensor (also represented as $S_{ij}$), s$^{-1}$ \\
$T$ & temperature, K \\
$\mathbf{T}^{(n)}$ & $n$--th tensor basis \\
$t$ & time, s \\
$t_s$ & $1/(\beta^* \omega)$; turbulent time scale, s \\
$T_r$ & $\left( 1+0.89(\gamma - 1)M_\infty^2 \right)T_\infty$; recovery temperature, K \\
$T_{0,e}$ & $T_e + \frac{1}{2 c_p}U_e^{2}$ ; total temperature at the boundary layer edge, K \\
$Tu$ & $100\sqrt{\frac{2}{3}\frac{k_\infty}{U_\infty^2}}$; turbulent intensity \\
$U_\infty$ & freestream velocity, m/s \\
$u$ & streamwise velocity component, m/s \\
$u_j$ & velocity in $x_j$ direction, m/s \\
$u^{+}$ & $u/u_\tau$, streamwise velocity component in inner wall units\\
$u_{\mathrm{V}}^{+}$ & $\int_0^{u^{+}} \frac{\left(\bar{\rho} / \bar{\rho}_w\right)^{1 / 2}}{\left(\bar{\mu} / \bar{\mu}_w\right)^{1 / 2}} \mathrm{~d} u^{+}$, Volpiani transformed mean velocity \\
$u_\tau$ & $\sqrt{\tau_w / \rho_w}$; friction velocity, m/s \\
$x$ & streamwise direction, m \\
$x_a$ & streamwise location of sampling wall--normal profiles, m \\
$x_i$ & streamwise location where $\delta$ matches inflow boundary-layer thickness of direct numerical simulation data, m \\
$\mathrm{Y}$ & observation data \\
$y$ & wall--normal direction, m \\
$y^{+}$ & $y \rho_w u_\tau / \mu_w$, non--dimensional distance in wall--normal direction  \\
$y_V^{+}$ & $\int_0^{y^{+}} \frac{\left(\bar{\rho} / \bar{\rho}_w\right)^{1 / 2}}{\left(\bar{\mu} / \bar{\mu}_w\right)^{3 / 2}} \mathrm{~d} y^{+}$, Volpiani transformed distance in wall--normal direction  \\
$\delta$ & boundary--layer thickness (based on 99\% of the freestream velocity), m \\
$\delta_i$ & boundary-layer thickness at inflow plane of direct numerical simulation data, mm \\
$\epsilon_h$ & discretization error \\
$\gamma$ & $c_p/c_v$, specific heat ratio \\
$\mathcal{H}$ & model operator \\
$\mu$ & dynamic viscosity, kg/(m$\cdot$s) \\
$\mu_t$ & eddy viscosity, kg/(m$\cdot$s) \\
$\rho$ & density, kg/m$^3$ \\
$\sigma_{ij}$ & viscous stress tensor, Pa \\
$\boldsymbol{\tau}$ & Reynolds stress tensor (also represented as $\tau_{ij}$), Pa \\
$\tau_w$ & wall shear stress, Pa \\
$\boldsymbol{\tau}^{\mathrm{dev}}$ & deviatoric part of the Reynolds stress tensor, Pa \\
$\omega$ & turbulent specific dissipation, s$^{-1}$ \\
$\Omega$ & $\frac{1}{2} \left( \frac{\partial u_i}{\partial x_j} - \frac{\partial u_j}{\partial x_i} \right)$; rotation rate tensor (also represented as $\Omega_{ij}$), s$^{-1}$ \\
$\theta$ & scalar invariants \\
$\boldsymbol{w}$ & neural network parameters \\
\multicolumn{2}{@{}l}{Subscripts}\\
$aw$ & adiabatic wall value \\
$e$ & edge condition \\
$\infty$ & freestream quantity  \\
$w$ & wall variable \\
\multicolumn{2}{@{}l}{Superscripts}\\
$l$ & iteration index during model training \\
$+$ & variable in inner wall units \\
\end{longtable*}}

\section{Introduction}
\lettrine{T}{urbulence} modeling is arguably the most crucial aspect of high Reynolds number hypersonic flow computations. Hypersonic flight conditions are characterized by extremely high kinetic energy in the flow, which dissipates as heat energy near the surface of the vehicle. The heating rates in the turbulent boundary layer can be several times higher than the laminar boundary layer. Furthermore, as a result of radiative cooling and internal heat transfer, the external surface temperatures of hypersonic flight vehicles are often considerably lower than the adiabatic wall temperature. Surface temperatures may remain below the adiabatic wall temperature during brief or prolonged flights, depending on the vehicle’s design (such as heat management systems and surface materials) and the flight trajectory (e.g., short reentry missions, quick experimental runs in wind tunnel testing, or sustained hypersonic flights). These cold wall conditions result in substantial thermal gradients between the surface and the fluid, which in turn modify the characteristics of near-wall turbulence. Similar cold wall conditions can also arise in wind tunnel experiments, where the short duration of the experiments prevents the test article from reaching higher temperatures. Therefore, the accurate modeling of cold--wall hypersonic turbulent boundary layers is of utmost importance for predicting surface heat flux and ensuring reliable thermal protection design for such vehicles.

The most commonly used turbulence models in Reynolds-averaged Navier-Stokes (RANS) simulations are typically calibrated for incompressible flow conditions. Extensive validation has been conducted for subsonic or moderately supersonic flows, where significant near--wall heat flux is absent. According to the Morkovin hypothesis~\cite{morkovin1962}, when variations in mean thermodynamic quantities are appropriately accounted for, incompressible turbulence models have demonstrated satisfactory performance in capturing mean velocity and mean temperature fields in compressible turbulent boundary layers. So et al.~\cite{so1998} further established that the Morkovin hypothesis extends to turbulence quantities, indicating a dynamic similarity between the near-wall turbulence characteristics of incompressible and compressible wall-bounded turbulent flows. However, the validity of this hypothesis diminishes as the Mach number increases into the hypersonic regime, and these turbulence models exhibit progressively poorer performance, particularly in cold-wall scenarios~\cite{rumsey2010}. To address these limitations, several compressibility corrections have been proposed. A comprehensive review and evaluation of turbulence models for hypersonic flows can be found in Ref.~\cite{roy2006pas}. Compressibility corrections for hypersonic boundary layer applications have also been analyzed in Ref.~\cite{rumsey2010} that emphasizes the existing need for improved turbulence models specifically tailored for hypersonic flows.

In cold--wall cases, direct numerical simulations (DNS) have revealed that the effect of wall cooling can significantly alter the near--wall turbulence structure~\cite{zhang2017M6}, which is not adequately accounted for by RANS models. At hypersonic flow conditions, RANS turbulence models have been shown to overestimate wall heat transfer and skin--friction by up to $30\%$~\cite{aiken2022RANS} when compared to DNS data~\cite{zhang2018dns, huang2020M11}. Moreover, as the wall cooling effect and freestream Mach number increase, RANS models tend to overestimate the peak temperature in the boundary layer by nearly $25\%$~\cite{aiken2022RANS}. In an effort to address these issues, a compressibility correction was proposed and analyzed for the $k$--$\omega$ model in cold-wall hypersonic flow cases~\cite{danis2022kOmegaCorr}. This compressibility correction demonstrated improvements in velocity and skin-friction estimation. However, its impact on the computation of boundary layer peak temperature and heat transfer was minimal. The proposed compressibility correction primarily focuses on enhancing the prediction of eddy viscosity to improve the estimation of wall heat flux as a second-order effect.

One notable source of uncertainty in turbulence modeling, particularly for hypersonic flows, is the assumption of a constant turbulent Prandtl number throughout the boundary layer and across different Mach numbers. This assumption has been identified as a source of uncertainty in various studies, particularly those related to hypersonic flows~\cite{rumsey2010, aiken2022RANS}. Researchers have addressed this problem by introducing a variable turbulent Prandtl number formulation based on a two-equation model for enthalpy variance and its dissipation rate~\cite{xiao2007prt}. They demonstrated improvements in heat transfer prediction for flows involving separated regions and shock--wave/boundary-layer interactions. Therefore, there is a strong case for incorporating a modeling approach that considers variable turbulent Prandtl number along with Reynolds Stresses in the development of turbulence models specifically tailored for hypersonic flows.

In the domain of data--driven turbulence modeling~\cite{duraisamy2019data}, machine learning methods have been predominantly employed over the past decade to systematically utilize data for the development of more robust and generalizable turbulence models~\cite{ling2016jfm, weatheritt2016jcp, wu2018prf, han2023equivariant}. Typically, these machine learning models have been trained using \textit{direct data} of closure terms (i.e. Reynolds stresses) obtained from high--fidelity simulations (e.g. DNS). However, the availability of such high--fidelity data is primarily limited to simple geometries or low Reynolds numbers. Furthermore, when trained turbulence models are coupled with RANS solvers, they often yield poor predictions of mean flow quantities~\cite{wu2019jfm}. This discrepancy arises due to inconsistencies between the training and prediction environments. The inconsistencies may arise due to features mismatch between training and prediction, ill--conditioning of RANS equations with explicitly trained closure models, and error accumulation over time in prediction environment~\cite{duraisamy2021prf}. Therefore, it is desirable to utilize \textit{indirect data}, which encompasses quantities like velocity, temperature, lift coefficients, etc., within a model--consistent learning framework. A model-consistent learning framework ensures that the turbulence model is trained in a manner consistent with its intended use in actual simulations (e.g., within a RANS solver). Thus, this training process involves RANS solvers and requires solving an inverse problem to minimize the discrepancy between the model output and high-fidelity data. Numerous approaches have been proposed in this regard, including adjoint-based methods~\cite{holland2019adjoint,carlos2021adjoint}, ensemble-based learning~\cite{carlos2021ensemble}, symbolic regression~\cite{saidi2022symbolic}, and the ensemble Kalman method~\cite{zhang2022jfm}. In terms of directly modeling turbulent heat flux closure, a model--consistent approach based on gene-expression programming (GEP) has been employed to learn varying turbulent Prandtl number for an incompressible flow case, although the eddy viscosity term was not learned in this approach~\cite{lav2021}.

In this work, a neural network--based turbulence model for hypersonic boundary layer flow over a cold wall is developed, using a model--consistent framework based on iterative ensemble Kalman method. 
Specifically, a linear eddy viscosity model along with a variable turbulent Prandtl number is learned to enhance the accuracy of wall heat flux predictions. Spatially sparse observation data of velocity and temperature are used for the training. To train the neural network (NN), the iterative ensemble Kalman method is employed, which has demonstrated superior efficiency compared to adjoint--based learning methods~\cite{zhang2022jfm}. Unlike the gradient-free GEP method, the ensemble Kalman method approximates the gradient and Hessian based on the statistics of the ensemble of simulation results, resulting in improved efficiency. Furthermore, in contrast to adjoint-based methods, the ensemble Kalman method requires significantly less implementation effort due to its non-intrusive nature.

This paper represents a preliminary effort in creating data--driven turbulence models for hypersonic flows, subsequently aiming to address a broader spectrum of flow scenarios beyond ZPG flow cases, such as those involving nonzero pressure gradients, shock-boundary layer interactions, and flow separation. To this end, the observation data utilized here for training is obtained from a recent DNS database for hypersonic boundary layer flows over a flat plate with zero--pressure gradient (ZPG). The database includes various flow cases with wall--to--recovery temperature ratios ($T_w/T_r$) ranging from 0.18 to 0.76 and Mach numbers between 6 and 14. The remainder of the paper is organized as follows. \S~\ref{sec:methodology} introduces the representation of closure terms for compressible flows and provides an overview of the employed training framework. \S~\ref{sec:flow_cases} presents and analyzes benchmark hypersonic flow cases, along with the corresponding grids generated for the respective RANS simulations. \S~\ref{sec:results} presents and discusses the training and testing results. Finally, \S~\ref{sec:conclusion} concludes the paper.

\section{Methodology}
\label{sec:methodology}
For compressible flows, the Favre--averaged equations for conservation of mass, momentum, and energy can be written as:
\begin{subequations}
\label{eqn:favre}
\begin{align}
& \frac{\partial \bar{\rho}}{\partial t}+\frac{\partial}{\partial x_j}\left(\bar{\rho} \hat{u}_j\right)=0 \\
& \frac{\partial\left(\bar{\rho} \hat{u}_i\right)}{\partial t}+\frac{\partial}{\partial x_j}\left(\hat{u}_j \bar{\rho} \hat{u}_i\right)=-\frac{\partial \bar{p}}{\partial x_i}+\frac{\partial \bar{\sigma}_{i j}}{\partial x_j}+\frac{\partial \tau_{i j}}{\partial x_j} \\
& \frac{\partial(\bar{\rho} \hat{E})}{\partial t}+\frac{\partial}{\partial x_j}\left(\hat{u}_j \bar{\rho} \hat{H}\right)=\frac{\partial (\bar{\sigma}_{i j} \hat{u}_i)}{\partial x_j} + \frac{\partial (\hat{u}_i \tau_{i j})}{\partial x_j} - \frac{\partial (\bar{q}_j)}{\partial x_j} - \frac{\partial (c_p \overline{\rho u_j^{\prime \prime} T^{\prime \prime}})}{\partial x_j} 
\end{align}
\end{subequations}
where the hat sign ( $\hat{ }$ ) over a variable represents the Favre--average~(density--weighted) quantity and the overbar ( $ \bar{ } $ ) denotes Reynolds--averaged quantity. 
The viscous stress tensor ($\bar{\sigma}_{ij}$) and laminar heat flux ($\bar{q}_j$) are described as:
\begin{equation}
\bar{\sigma}_{i j} \approx 2 \hat{\mu}\left(\hat{S}_{i j}-\frac{1}{3} \frac{\partial \hat{u}_k}{\partial x_k} \delta_{i j}\right) ,
\qquad \qquad 
\bar{q}_j=-\overline{k_T \partial T / \partial x_j} \approx-\frac{c_p \hat{\mu}}{P_r} \frac{\partial \hat{T}}{\partial x_j}
\end{equation}
where thermal conductivity ($k_T$) is approximated using dynamic viscosity ($\mu$) and Prandtl number (Pr). 
In the above Favre--averaged equations (Eq.~\ref{eqn:favre}), the Reynolds stress ($\tau_{ij}$) and turbulent heat flux ($q_j^{(t)} = c_p \overline{\rho u_j^{\prime \prime} T^{\prime \prime}}$) terms need to be modeled. 

\subsection{Modeling of closure terms}
Reynolds stress term ($\boldsymbol{\tau}=\tau_{ij}$) can be decomposed into a deviatoric part ($\boldsymbol{\tau}^{\mathrm{dev}}$) and a spherical (or dilatational) part, as:
\begin{equation}
\boldsymbol{\tau} = \boldsymbol{\tau}^{\mathrm{dev}} - \frac{2}{3} \bar{\rho} k \mathbf{I}
\end{equation}
where $k$ is the turbulent kinetic energy and $\mathbf{I}$ represents the identity matrix. Based on the general effective--viscosity model~\cite{pope1975}, the deviatoric part ($\boldsymbol{\tau}^{\mathrm{dev}}$) of Reynolds stresses can be formulated as a function of only the normalized strain rate tensor $\hat{\mathbf{S}}$ and the rotation rate tensor $\hat{\mathbf{\Omega}}$. Normalization of both these tensors is defined and discussed in Sec.~\ref{sec:methodology}B.

The most general form of $\boldsymbol{\tau}^{\mathrm{dev}}$ can be expressed as a linear combination of isotropic basis tensors ($\mathbf{T}$):
\begin{equation}
\boldsymbol{\tau}^{\mathrm{dev}} = 2 \bar{\rho} k \sum_{n=1}^{10} g^{(n)} \mathbf{T}^{(n)}
\label{eqn:basis_rep}
\end{equation}
where scalar coefficients $g^{(n)}$ may be functions of independent invariants of $\hat{\mathbf{S}}$ and $\hat{\mathbf{\Omega}}$\cite{pope1975}. Owing to the Cayley–Hamilton theorem, there is only a finite number of independent basis tensors and invariants that can be formed from $\hat{\mathbf{S}}$ and $\hat{\mathbf{\Omega}}$. Detailed derivation of these independent basis tensors can be found in Ref.\cite{pope1975}, and the linear and quadratic basis tensors are listed here:
\begin{equation}
\mathbf{T}^{(1)}=\hat{\mathbf{S}}-\frac{1}{3} \{\hat{\mathbf{S}}\}, \qquad 
\mathbf{T}^{(2)}=\hat{\mathbf{S}} \hat{\mathbf{\Omega}}-\hat{\mathbf{\Omega}} \hat{\mathbf{S}}, \qquad 
\mathbf{T}^{(3)}=\hat{\mathbf{S}}^2-\frac{1}{3} \left\{\hat{\mathbf{S}}^2\right\}, \qquad 
\mathbf{T}^{(4)}=\hat{\mathbf{\Omega}}^2-\frac{1}{3} \left\{\hat{\mathbf{\Omega}}^2\right\}
\end{equation}
where the curly bracket $\{\cdot \}$ denotes the trace of the matrix.

The representation of the scalar coefficients ($g^{(n)}$) can be determined as a function of the finite set of independent invariants of $\hat{\mathbf{S}}$ and $\hat{\mathbf{\Omega}}$. The derivation of this basis of invariants has been discussed in Ref.~\cite{ling2016jcp}. Using this basis of invariants embeds the Galilean invariance in the model representation.
For two--dimensional compressible flows, there are only three non-zero independent invariants\cite{ling2016jcp, wallin2000}:
\begin{equation}
\theta_1=\{\hat{\mathbf{S}}\}, \quad \theta_2=\left\{\hat{\mathbf{S}}^2\right\}, \quad \theta_3=\left\{\hat{\mathbf{\Omega}}^2\right\}
\label{eqn:theta}
\end{equation}
In the case of compressible flows, the consideration of a non-zero trace of the strain rate is necessary~\cite{wallin2000}, leading to the first invariant $\theta_1$. 
The primary challenge in deriving an explicit relation for the deviatoric Reynolds stress ($\boldsymbol{\tau}^{\mathrm{dev}}$) in Eq.~\ref{eqn:basis_rep} lies in determining the scalar coefficients ($g^{(n)}$). Shih et al.~\cite{shih1993algebraic} derived these coefficients algebraically for a selected set of basic flows. In this work, the functional representation of the scalar coefficients, based on invariants, is modeled using a neural network, an approach that was proposed in Ref.~\cite{ling2016jfm}.

For this work, which focuses on attached, zero-pressure-gradient boundary layer flows, we consider only a linear representation term of deviatoric Reynolds stress, akin to linear eddy viscosity model following the Boussinesq hypothesis ($\boldsymbol{\tau}^{\mathrm{dev}}=2 \mu_t (\mathbf{S} - (1/3)\{\mathbf{S}\})$)~\cite{boussinesq1987}. Consequently, from the general expression in Eq.~\ref{eqn:basis_rep} of effective--viscosity model~\cite{pope1975}, the linear turbulent eddy viscosity ($\mu_t$) is estimated using only the first term 
($\boldsymbol{\tau}^{\mathrm{dev}} = 2 \bar{\rho} k g^{(1)} T^{(1)}$)
as:
\begin{equation}
\mu_{\mathrm{t}}=-g^{(1)} \bar{\rho} k t_s
\label{eqn:nut}
\end{equation}
where $g^{(1)}$ is the scalar coefficient of the first tensor basis. The negative sign on the right-hand side ensures that the negative values of $g^{(1)}$ yield a final positive value for $\mu_{\mathrm{t}}$. The turbulence time scale, given by $t_s=1/(\beta^* \omega)$, establishes a relationship between the mean strain rate tensor $\mathbf{S}$ and its normalized counterpart $\hat{\mathbf{S}} = t_s \mathbf{S}$. $\beta^*$ is a model constant, and the specific turbulence dissipation rate ($\omega$) is obtained by solving the respective transport equation given by the $k$-$\omega$ turbulence model.

Second closure term is the turbulence heat flux 
($q_j^{(t)}$), for which Reynolds analogy is most commonly used to model the term as:
\begin{equation}
q_j^{(t)} = c_p \overline{\rho u_j^{\prime \prime} T^{\prime \prime}} \approx-\frac{c_p \mu_t}{P r_t} \frac{\partial \hat{T}}{\partial x_j}
\label{eqn:turbheatFlux}
\end{equation}
where the turbulent Prandtl number ($Pr_{t}$) is typically assumed to have a constant value. However, in this work, we address the uncertainty associated with this assumption of constant value by learning a variable $Pr_{t}$.

\subsection{Machine Learning Formulation}
Based on the aforementioned representations of the closure terms for Reynolds stress and turbulent heat flux, the closure modeling involves the learning of a scalar coefficient $g^{(1)}$ and a turbulent Prandtl number $Pr_{t}$.
Although a functional form is assumed for the closure terms (i.e., a linear representation of the deviatoric Reynolds stress and the Reynolds analogy for the turbulence heat flux), the current approach is akin to machine learning-based closure modeling~\cite{ling2016jfm, carlos2021adjoint, zhang2022jfm}. This method is not tied to a specific turbulence model, offering greater flexibility and potential for accuracy. However, it may require significant training data and computational effort, while posing challenges in physical interpretability and generalization. In contrast, the model augmentation approach~\cite{Parish2016, Singh2017, holland2019adjoint} leverages existing turbulence models by integrating data-driven insights through a correction term, which limits its flexibility and accuracy potential.

A neural network--based closure model is employed to predict two output quantities ($g^{(1)}$ and $Pr_{t}$) as a function of seven input features. The input features set is given as:
$$ \left[ \{\hat{\mathbf{S}}\}, \qquad \left\{\hat{\mathbf{S}}^2\right\}, \qquad \left\{\hat{\mathbf{\Omega}}^2\right\}, \qquad \frac{\frac{\partial T}{\partial n}*l_t}{\bar{T}}, \qquad \frac{\nu_t}{100\nu + \nu_t}, \qquad  \tanh\left( d_{wall}\frac{\sqrt{k}}{100\nu} \right), \qquad \frac{T_w-T_\delta}{T_r-T_\delta} \right] $$
The first three input features are the independent scalar invariants of $\hat{\mathbf{S}}$ and $\hat{\mathbf{\Omega}}$, as given in Eq.~\ref{eqn:theta}. Here, strain rate tensor $\hat{\mathbf{S}}=\mathbf{S}/\left(\|\mathbf{S}\|+1 / t_s \right)$ and rotation rate tensor $\hat{\mathbf{\Omega}}=\mathbf{\Omega}/\left(\|\mathbf{\Omega}\|+1 / t_s \right)$ are normalized using a local quantity of turbulence time scale $t_s$. 
The fourth feature is based on the wall--normal temperature gradient, normalized by the local cell--averaged temperature $\bar{T}$ and the local turbulent length scale $l_t = \sqrt{k} / (\beta^* \omega)$. Along with the magnitude, the sign of the temperature gradient is particularly important for calculating the wall heat flux. Features based on the temperature gradient have also been used previously to develop heat flux closures for incompressible flows~\cite{lav2021, milani2018}. The fifth feature is based on the turbulent viscosity ratio, formulated in such a way as to constrain the feature within the range [−1, 1] and reduce the probability of the denominator approaching zero.
The sixth feature is the wall distance Reynolds number with a smooth cutoff. The final input feature is the dimensionless wall temperature, which characterizes the effects of wall temperature in diabatic flow, where $T_w$, $T_r$, and $T_\delta$ represent the wall temperature, the recovery temperature at the wall, and the temperature at the edge of the boundary layer, respectively. This feature is also referred to as the \textit{diabatic parameter} and provides a unique parameter for determining the mean temperature-velocity relationship in flows under different wall temperature conditions~\cite{zhang2014GRA}. Although the determination of this feature requires nonlocal values at the edge of the boundary layer, the inclusion of this feature provides a basis for distinguishing flows with different wall-to-recovery temperature ratios.

Among the input features mentioned above, the first six are normalized using local flow quantities to yield values within the range [−1,1]. Such local normalization helps in avoiding skewed distributions and improves the convergence behavior during training, as also observed in earlier studies~\cite{ling2015pof, wu2018prf, liu2023transonic}. Local normalization also offers greater possibility to generalize across different flow configurations as compared to global normalization~\cite{duraisamy2021prf}. 
Note that for hypersonic flows over a flat plate, $\theta_1$ has non-zero values at the leading edge due to compressibility effects. Beyond the leading edge, the value of $\theta_1$ is uniformly very small (almost zero). The presence of such an almost zero input feature will not affect the modeling, as it will be ignored during the training of the NN--based closure model.

\subsection{Ensemble Kalman method for training}
To learn the NN--based closure model, we employ the ensemble Kalman method to infer the parameters $\boldsymbol{w}$, which include all the weights and biases of the fully connected neural network. The ensemble Kalman method is a data assimilation technique that finds application in a diverse range of fields and disciplines~\cite{schneider2022jcp, zhang2022jfm, zhang2022aiaa, zhou2023inference}. It effectively combines the information from an ensemble of model simulations and the available observational data to provide a statistical inference of the system's state. In the current research work, the system's state corresponds to the parameters $\boldsymbol{w}$ of the NN--based closure model. For the iterative process of training these parameters, the corresponding cost function is formulated as:
\begin{equation}
J=\left\|\boldsymbol{w}^{l+1}-\boldsymbol{w}^l\right\|_{\mathrm{P}}+\left\|\mathrm{Y} - \mathcal{H}\left[\boldsymbol{w}^{l+1}\right]\right\|_{\mathrm{R}}
\label{eqn:cost_function}
\end{equation}
where $l$ is the iteration index, and $\left\| \cdot \right\|_A$ represents the weighted norm (defined as $\|\boldsymbol{v}\|_A^2=\boldsymbol{v}^{\mathrm{T}} A^{-1} \boldsymbol{v}$) for a vector $\boldsymbol{v}$ with weight matrix $A$. $\mathrm{P}$ represents the model error covariance matrix, reflecting the uncertainties associated with the model parameters, while $\mathrm{R}$ denotes the observation error covariance matrix for the observation data $\mathrm{Y}$. During each iteration, the parameters $\boldsymbol{w}$ are mapped to quantities in observation space (e.g., velocity and temperature) by the operator $\mathcal{H}$. Hence, $\mathcal{H}$ denotes the process of predicting closure variables, forward propagation of RANS equations and post-processing, and the term $\mathcal{H}\left[\boldsymbol{w}\right]$ represents the predicted observed quantities. 
The first term in Eq.~\ref{eqn:cost_function} provides regularization for the updated parameters $\boldsymbol{w}$ by penalizing large changes, while the second term penalizes the difference between the observational data $\mathrm{Y}$ and the model prediction $\mathcal{H}[\boldsymbol{w}]$.

\begin{figure} [hbt!]
\centering
\includegraphics[width=0.7\textwidth]{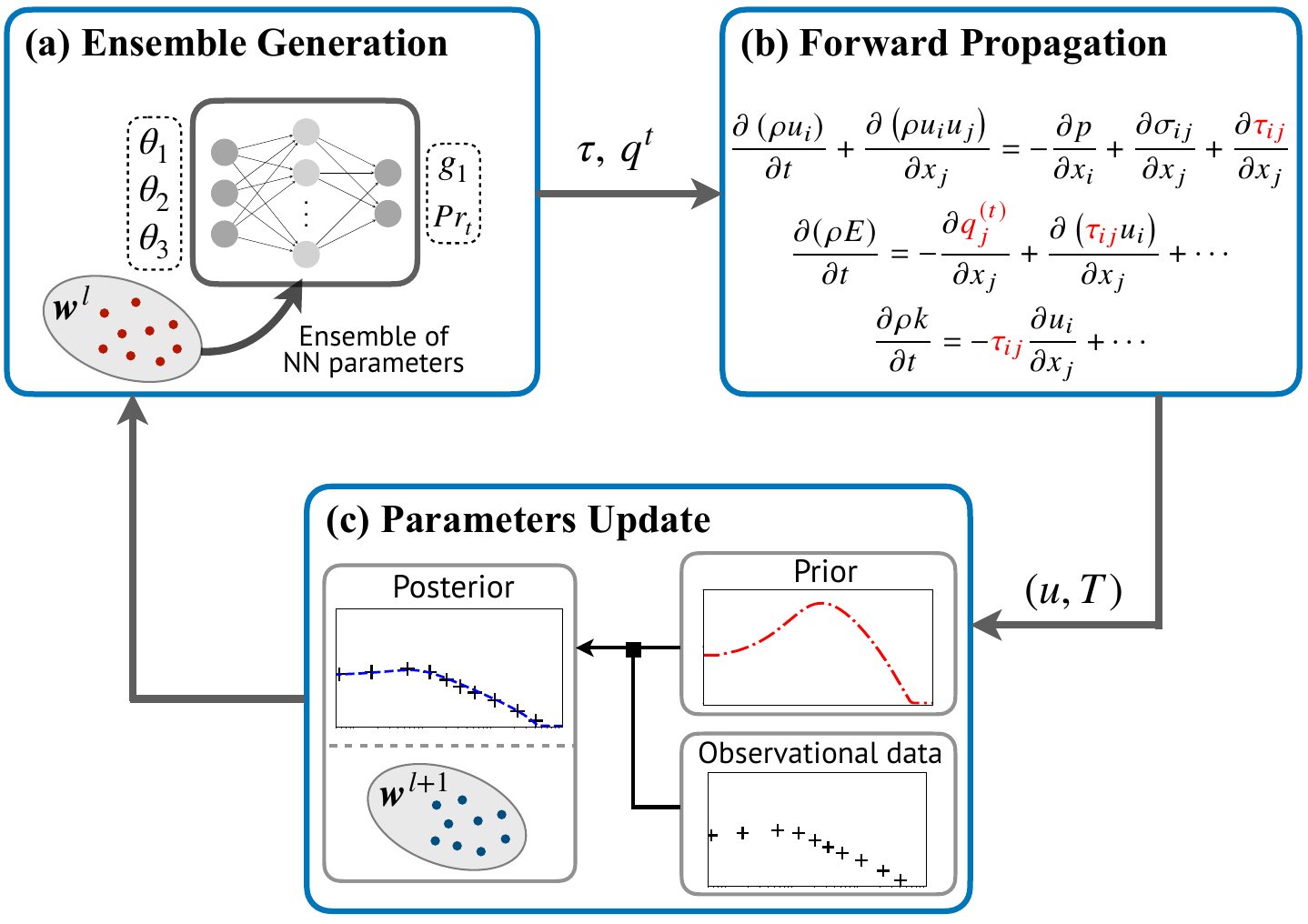}
\caption{Schematic of training NN--based turbulence model using ensemble Kalman method.}
\label{fig:schematic}
\end{figure}

In the ensemble Kalman method, the update scheme for the parameters $\boldsymbol{w}$ during each iteration is formulated as:
\begin{subequations}
\label{eqn:update}
\begin{equation}
\boldsymbol{w}_m^{l+1}=\boldsymbol{w}_m^l + \mathrm{K} \left( \mathrm{Y}_m-\mathcal{H}\left[\boldsymbol{w}_m^l\right]\right)
\end{equation} 
\begin{equation}
\text{with} \quad \mathrm{K} = \mathrm{S_w} \mathrm{S_y^T}\left( \mathrm{S_y} \mathrm{S_y^T} + \mathrm{R}\right)^{-1}
\end{equation}
\end{subequations}
where $m$ is the sample index. The Kalman gain matrix $\mathrm{K}$ has been written in an equivalent form to the commonly used form of $\mathrm{K} = \mathrm{P} \mathrm{H}^{\mathrm{T}}\left( \mathrm{H} \mathrm{P} \mathrm{H}^{\mathrm{T}} + \mathrm{R}\right)^{-1}$. Equivalence is established by writing the model error covariance matrix $\mathrm{P}$ and other associated quantities in terms of the square root matrix $\mathrm{S_w}$ and its projection $\mathrm{S_y}$ to the observation space, i.e.
$\mathrm{P} = \mathrm{S_w} \mathrm{S_w^T}$ and $\mathrm{S_y} = \mathrm{H}\mathrm{S_w}$. Here, $\mathrm{H}$ is the gradient of the observation operator $\mathcal{H}$ with respect to the model parameters $\boldsymbol{w}$. Consequently, the equivalence of the cross-covariance term and the projection of $\mathrm{P}$ to the observation space are: 
\begin{equation*}
\mathrm{P}\mathrm{H^T}=\mathrm{S_w}\mathrm{S_y^T} \quad \text{and} \quad \mathrm{H}\mathrm{P}\mathrm{H^T}=\mathrm{S_y}\mathrm{S_y^T}
\end{equation*}
To estimate the Kalman gain matrix $\mathrm{K}$ from the ensemble using Eq.~\ref{eqn:update}b, the square root matrices at iteration step $l$ are computed as follows:
\begin{subequations}
\label{eqn:sqrt_matrices}
\begin{equation}
\mathrm{S}_w^l=\frac{1}{\sqrt{N_e-1}}\left[\boldsymbol{w}_1^l-\overline{\boldsymbol{w}}^l, \boldsymbol{w}_2^l-\overline{\boldsymbol{w}}^l, \ldots, \boldsymbol{w}_{N_e}^l-\overline{\boldsymbol{w}}^l\right]
\end{equation} 
\begin{equation}
\mathrm{S}_y^l=\frac{1}{\sqrt{N_e-1}}\left[\mathcal{H}\left[w_1^l\right]-\mathcal{H}\left[\bar{w}^l\right], \mathcal{H}\left[\boldsymbol{w}_2^l\right]-\mathcal{H}\left[\bar{w}^l\right], \ldots, \mathcal{H}\left[\boldsymbol{w}_{N_e}^l\right]-\mathcal{H}\left[\bar{w}^l\right]\right]
\end{equation}
\end{subequations}
where $N_e$ is the sample size, and $\overline{\boldsymbol{w}}^l$ represents the ensemble mean of the parameters $\boldsymbol{w}$ at iteration $l$. For a detailed derivation of the mathematical relations above, and the relation between the approximated derivatives of the cost function and the Kalman gain matrix, we recommend referring to Ref.~\cite{zhang2022jfm}. The implementation of the iterative training process using the ensemble Kalman method has been summarized in Algortihm~\ref{EnKF_algo}. The schematic of the training framework for the current application of hypersonic flows is depicted in Fig.~\ref{fig:schematic}.

\begin{algorithm}
    \caption{Model Training with Ensemble Kalman method}\label{EnKF_algo}
    \begin{algorithmic}
        \STATE Pre-training of parameters $\boldsymbol{w}$
        \STATE Generate state ensemble (parameters $[\boldsymbol{w_1, w_2, \dots w_{N_e}} ]$)
        \FOR{iter = $1$ to $N$}
            \STATE Extracting input features from flow field
            \STATE Predict outputs ($g_1$ and $Pr_t$) corresponding to each sample state ($\boldsymbol{w}_m$)
            \STATE Forward propagate RANS equations to update ensemble of flow fields
            \STATE Obtain observation quantities ($U$ and $T$) at streamwise location $x_a$ 
            \STATE Estimate Kalman gain matrix using Eq.~\ref{eqn:update}b 
            \STATE Update state ensemble using Eq.~\ref{eqn:update}a
        \ENDFOR
    \end{algorithmic}
\end{algorithm}

Traditional initialization (such as random initialization or Xavier initialization) of the NN parameters $\boldsymbol{w}$ can lead to the prediction of non--physical values for the closure variables, which in turn can cause the flow solver to diverge. Therefore, the NN parameters are pre-trained using the output values of the baseline model, specifically $g^{(1)} = -0.09$ and $Pr_t = 1.0$. 
Note that generally, a constant value from the range of 0.85 to 1.0 is assumed for $Pr_t$ in RANS turbulence models. While a particular choice of $Pr_t$ within this range has a small effect on the RANS simulation results, this choice of $Pr_t = 1.0$ does not affect the results of the trained NN-based closure model, as will be demonstrated in the results.

The ensemble of NN parameters $\boldsymbol{w}$ is generated based on a normal distribution prescribed using absolute and relative standard deviations. In each subsequent iteration, this ensemble of NN parameters $\boldsymbol{w}$ is updated based on Eq.\ref{eqn:update}a. For the first iteration, input features are extracted from the baseline flow field, while for subsequent iterations, input features for each sample are extracted from the flow field updated during the previous iteration. Using each sample $\boldsymbol{w}_m$, the input features are mapped to closure variables ($g^{(1)}$ and $Pr_t$). Each sample $\boldsymbol{w}_m$ represents a different NN turbulence closure model, resulting in an ensemble of predicted Reynolds stress and turbulent heat flux fields. To compute the updated mean flow quantities, the RANS equations are forward-propagated using each of the predicted Reynolds stress and turbulent heat flux fields. During post-processing, mean flow quantities (e.g., velocity and temperature) are extracted at specified streamwise location along the flat plate for each sample in the ensemble of RANS simulation results. Based on the misfit between this ensemble of mean flow quantities in observation space and the observation data from DNS, the updated parameters $\boldsymbol{w}$ are statistically inferred based on Eq.~\ref{eqn:update}.

The DAFI code~\cite{carlos2021dafi} is employed to implement the ensemble--based training. The NN--based closure model is represented by a fully connected neural network with 10 hidden layers, each consisting of 10 neurons. 
This specific neural network size is derived from the original work of Ling et al.~\cite{ling2016jfm}. The same architecture has also been used in other turbulence modeling studies~\cite{carlos2021adjoint, carlos2021ensemble, zhang2022jfm}. Notably, the sensitivity study conducted by Zhang et al.~\cite{zhang2022jfm} indicates that the size of the neural network architecture does not significantly impact training accuracy.
The rectified linear unit (ReLU) activation function is used to introduce non--linearity in the hidden layers, while linear activation is applied to the output layer. RANS simulations are performed using OpenFOAM, an open--source CFD tool based on the finite volume method. The simulations utilize the built--in explicit density-based compressible flow solver \textit{rhoCentralFoam}, which incorporates the second order central--upwind flux scheme proposed by Kurganov and Tadmor~\cite{kurganov2000new}. Second order schemes are used to approximate the gradient, divergence and Laplacian terms, and the van Leer limiter is used to approximate the solution variables at cell faces. All simulations are conducted with double precision.

\section{Hypersonic Flow Cases}
\label{sec:flow_cases}
The NN--based turbulence modeling approach has been analyzed using a DNS database from recently conducted hypersonic flow simulations~\cite{zhang2018dns, huang2020M11}. The simulations correspond to zero--pressure gradient flat plate turbulent boundary layer flows under cold--wall and near adiabatic conditions. These benchmark flow cases are summarized in Table~\ref{tab:flow_cases}, where freestream conditions and wall temperature for each case are provided. The listed cases cover a wide range of freestream Mach numbers and wall--to--recovery temperature ratios.  
All the flow cases fall within the perfect gas regime. The working fluid is air, except for the case M8Tw048, in which the working fluid is nitrogen.
\begin{table}[htbp]
\centering
\caption{Freestream conditions and wall temperatures for hypersonic flow cases in the DNS database \label{tab:flow_cases}}
\begin{tabular}{l r r r r r r r c} \hline
Case & $M_\infty$ & $U_\infty, \text{m/s}$ & $\rho_\infty, \text{kg/m}^3$ & $T_\infty, \text{K}$ & $T_w, \text{K}$ & $T_w/T_r$ & $\delta_i, \text{mm}$ & $(x_a-x_i)/\delta_i$ \\
\hline \hline
M6Tw025 & 5.84 & 869.1 & 0.044 & 55.2 & 97.5 & 0.25 & 1.3 & 88.6  \\ \hline
M6Tw076 & 5.86 & 870.1 & 0.043 & 55.0 & 300.0 & 0.76 & 13.8 & 54.1  \\ \hline
M8Tw048 & 7.87 & 1155.1 & 0.026 & 51.8 & 298.0 & 0.48 & 20.0 & 56.9  \\ \hline
M11Tw020 & 10.90 & 1778.4 & 0.103 & 66.5 & 300.0 & 0.20 & 3.8 & 127.0  \\ \hline
M14Tw018 & 13.64 & 1882.2 & 0.017 & 47.4 & 300.0 & 0.18 & 18.8 & 199.3  \\ \hline
\end{tabular}
\end{table}

\begin{figure}
\centering
\includegraphics[width=0.7\textwidth]{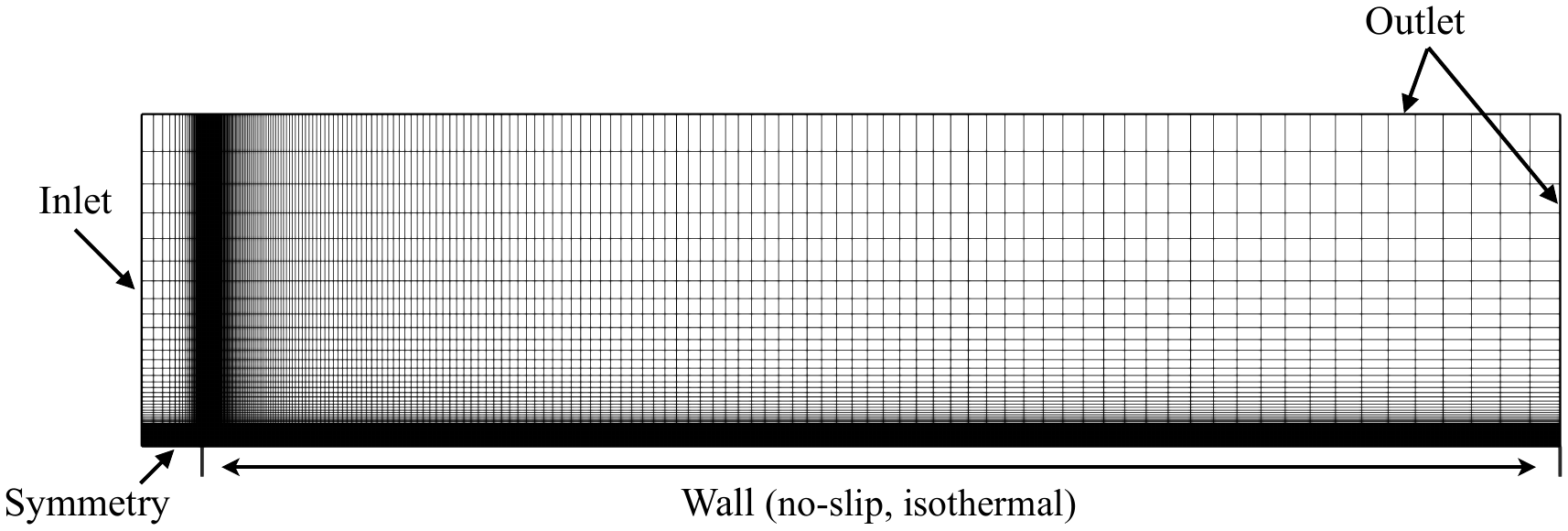} \qquad \qquad
\caption{Computational grid for the flat plate, with specified boundary conditions illustrated.}
\label{fig:grid_schematic}
\end{figure}

For each case, a rectangular domain grid is used, with a relatively shorter region upstream of the flat plate where a symmetry boundary condition is applied at $y=0$. 
The M6Tw076 grid is shown in Fig.~\ref{fig:grid_schematic} as an example to illustrate the specified boundary conditions.
The domain size in the wall--normal direction is set to ensure that the leading edge shock wave does not interact with the top boundary. In the streamwise direction, the flat plate is extended appropriately downstream of the sampling location. 

The sampling location ($x_a$) corresponds to the streamwise location along the flat plate where wall--normal profiles are provided in the DNS database~\cite{larcDNSdatabase, huang2022dnscases}. From the RANS simulation results of each case, $x_a$ is determined by first locating the position $x_i$ along the flat plate where the boundary layer thickness equals the inflow boundary layer thickness $\delta_i$ of the DNS data. Then, the sampling location $x_a$ is determined using the values of $\left(x_a-x_i\right) / \delta_i$ as specified in Table~\ref{tab:flow_cases}. A similar method for determining the sampling locations has also been employed in Ref.~\cite{danis2022kOmegaCorr} to compare RANS simulation results with the DNS database. The Reynolds number based on friction velocity and wall viscosity ($Re_{\tau}$) has also been used in Ref.~\cite{aiken2022RANS} to determine the sampling location for comparing RANS simulation results with DNS data. However, since $Re_{\tau}$ itself depends on the turbulence model, which is modified in each training iteration, we consider the sampling location with respect to the streamwise location $x_a$ here. Further investigation in this regard will be conducted in future work.

For each case, four levels of grids (ultra--fine, fine, medium, and coarse) are generated by systematically refining the meshes with a refinement factor of 2. The details of the medium-level grid for each case are provided in Table~\ref{tab:mesh_details}. 
\begin{table}[htbp]
\centering
\caption{Domain size and resolution of the medium level grid for RANS simulations. $\Delta y^+_{\text{a}}$ refers to the $y^+$ height of the first cell off the wall at the sampling location.
\label{tab:mesh_details}}
\begin{tabular}{l r r r c} \hline
Case & $L_x \times L_y, \text{m}$ & $x\text{--range, m}$ & $N_x \times N_y$ & $\Delta y^+_{\text{a}}$ \\
\hline \hline
M6Tw025 & $0.27 \times 0.10$ & ($-0.04, 0.23$) & $250 \times 250$ & $0.3$ \\ \hline
M6Tw076 & $2.35 \times 0.55$ & ($-0.10, 2.25$) & $520 \times 300$ & $0.3$ \\ \hline
M8Tw048 & $3.43 \times 0.80$ & ($-0.10, 3.33$) & $370 \times 260$ & $0.3$ \\ \hline
M11Tw020 & $1.55 \times 0.10$ & ($-0.05, 1.50$) & $760 \times 400$ & $0.3$  \\ \hline
M14Tw018 & $5.60 \times 0.70$ & ($-0.10, 5.50$) & $520 \times 350$ & $0.3$  \\ \hline
\end{tabular}
\end{table}

To assess the adequacy of the given RANS simulation results for their intended use, the numerical uncertainty associated with the discretization error at each mesh level is calculated. The discretization error depends on the chosen grid resolution, grid quality, and the numerical scheme~\cite{oberkampf2010book}. To evaluate the discretization error, the exact solution is approximated based on the results of the \textit{two finest grids} using the Richardson extrapolation (RE) method~\cite{roache1994RE, roy2005jcp}. 
The generalised form of RE is given by:
\begin{equation}
f_{\mathrm{RE}}=f_1 + \frac{f_1-f_2}{r^p-1} 
\label{eqn:RE}
\end{equation}
where $f$ is a generic solution variable considered as the quantity of interest. Among the two finest grids, $f_1$ corresponds to the finer one. $r$ represents the refinement factor of the grid, which is 2.0 in this work. $p$ represents the formal order of accuracy of the numerical schemes employed. Although second--order numerical schemes are used for simulations, it is not clear if the formal order of accuracy for this problem should be 1.0 or 2.0, as the flow contains a non--grid--aligned shock wave. Grid convergence error estimates in hypersonic flows, even in two-dimensional domains, are complicated by the presence of shock waves, which tend to reduce the spatial order of accuracy to first order, regardless of the nominal order of the spatial discretization scheme~\cite{carpenter1999, roy2003grid}. Hence, we adopt the conservative value of $p=1.0$.

In each case, the wall heat flux ($q_w$) at the sampling location ($x_a$) is considered as the quantity of interest. The discretization error ($\overline{\varepsilon}_h$) is evaluated by calculating the difference between the estimated exact solution ($f_{\mathrm{RE}}$) and the RANS solution. This error estimate is then converted to numerical uncertainty by considering the magnitude of the error estimate, along with an additional factor of safety ($F_s$), as follows:
\begin{equation}
\textit{Numerical Uncertainty} = F_s|\overline{\varepsilon}_h| = F_s|f_k - f_{RE}|
\end{equation}
where $f_k$ represents the value of the wall heat flux ($q_w$) at the sampling location ($x_a$) for the specific grid for which numerical uncertainty is being determined. The computed numerical solutions at different grid levels don't suggest to be in the asymptotic grid convergence range. The asymptotic range pertains to the range of discretization sizes where the lowest--order terms in the truncation error dominate. It is numerically established by checking whether the observed order of accuracy from numerical solutions of at least three grid levels matches the formal order~\cite{roy2005jcp}. Based on numerical solutions using four levels of systematically refined grids for each flow case, there is either a complete lack of asymptotic behavior for some flow cases, or the observed order of accuracy does not match the formal order of accuracy. Therefore, a conservative value of $F_s=3.0$ is used.

Numerical uncertainties, expressed as percentages of the estimated exact solutions, are presented in Table~\ref{tab:num_uncertainty} for the three coarsest grids of each flow case. 
The large jump between the numerical uncertainty values of coarse and medium grid levels can possibly be due to a large difference in laminar to turbulent transition locations. Further investigation is required to eliminate these disparities, potentially by tweaking the freestream turbulence quantities. Wall quantities of interest are significantly influenced by the location at which the turbulence model transitions from laminar to turbulent flow. This transition location varies across different grid levels. Increasing the freestream turbulence intensity may cause the transition to turbulent flow to occur earlier at all grid levels.
The provided values in Table~\ref{tab:num_uncertainty} suggest that the medium grid for each case offers a reasonable trade-off between the reliability and computational efficiency of the RANS simulations.
\begin{table}[htbp]
\centering
\caption{Numerical uncertainties for systematically refined meshes of each flow case, determined based on discretization error estimate and a factor of safety. \label{tab:num_uncertainty}}
\begin{tabular}{l c c c} \hline
Case & Coarse & Medium & Fine \\
\hline \hline
M6Tw025 & 17.96\% & 1.69\% & 1.01\% \\ \hline 
M6Tw076 & 43.18\% & 0.43\% & 0.67\% \\ \hline 
M8Tw048 & 23.07\% & 1.13\% & 1.87\% \\ \hline 
M11Tw020 & 13.79\% & 0.39\% & 0.20\%  \\ \hline 
M14Tw018 & 20.50\% & 4.91\% & 2.57\%  \\ \hline 
\end{tabular}
\end{table}

\section{Results}
\label{sec:results}
The ensemble--based training framework, which aims to learn a data--driven turbulence model for hypersonic flows, has been analyzed using two training setups: 
\begin{enumerate}
\item \textit{Single training case} (NN--1): Observation data from one flow case M6Tw025 are used for training 
\item \textit{Joint training case} (NN--2): Observation data from two flow cases, M6Tw076 and M14Tw018, are simultaneously used for training
\end{enumerate}
The trained turbulence models are subsequently tested on the remaining flow cases from Table~\ref{tab:flow_cases} for each of the training setups. Sparse observation data points of the wall--normal profiles for streamwise velocity ($u$) and temperature ($T$) within the boundary layer region at the sampling location are used for training. 
Observation data points outside the boundary layer region are excluded manually in this work, however this could be automated using the available turbulence functions that identify the boundary region (e.g. from Menter's SST model). 
The sparse observation points are extracted from the DNS database, with a higher concentration of points closer to the wall. This is achieved by systematically skipping an increasing number of points in the DNS database as we move away from the wall. 
This choice of sampling points avoids a high clustering of points towards the edge of the BL and results in roughly evenly spaced data locations when 
y+ is plotted on a log scale.
Moreover, during training, near--wall data points are weighted higher to place greater emphasis on improving the primary quantities of interest, i.e., wall heat flux and skin friction. This higher weighting is implemented by scaling the diagonal elements of the observation error covariance matrix $\mathrm{R}$ using factors within the range of [0.1, 1]. Scaling a diagonal element of $\mathrm{R}$ with a higher value corresponds to lower confidence in the respective observation data. Here, 0.1 corresponds to points at the wall, and 1.0 corresponds to points at the boundary layer edge, with the scaling values varying logarithmically from the wall to the boundary layer edge.

With these settings in place, the turbulence model training is conducted using two different training setups. The stopping criterion for training is based on the convergence of the cumulative misfit between observation data from DNS and that computed using the NN--based closure model.
The cumulative misfit is mathematically determined as:
$$ \mathcal{L} = \sqrt{\sum_{i=1}^m \sum_{j=1}^n \left|\mathrm{Y}_j - \mathrm{\hat{Y}}_j\right|_i^2} $$
for $m$ number of samples in the ensemble and $n$ observation data points of the wall–normal profiles of mean velocity and temperature for each sample. $\mathrm{Y}$ and $\mathrm{\hat{Y}}$ correspond to the observation data from DNS and  that computed using the NN–based closure model, respectively. 
Convergence is considered to be achieved when the absolute change in $\mathcal{L}$ for atleast five consecutive iterations is less than $1\%$.
For each of these training setups, the results for the respective training and test cases are compared with the DNS database based on flow quantities, including velocity and temperature, as well as wall heat transfer and skin friction coefficients. 
The van Driest transformation~\cite{vanDriest1951} is commonly applied to normalize the mean velocity profile in a compressible boundary layer; however, its effectiveness was found to deteriorate under hypersonic cold--wall conditions~\cite{duan2010dns, zhang2018dns}. Therefore, the Volpiani transformation~\cite{volpiani2020transformation} is used for the normalization of mean velocity and wall distance, as it has demonstrated an improved overall collapse of the mean velocity profile for hypersonic cold--wall boundary layers to the incompressible law of the wall.
The heat transfer coefficient is defined in two forms, and the appropriate choice between these two forms depends on $T_w/T_r$. For calorically perfect gases, the \textit{adiabatic wall} and \textit{edge} heat transfer coefficients are given as
\begin{equation}
C_{h,aw} = \frac{q_w}{\rho_{\infty} c_p U_{\infty} \left( T_r - T_w \right)}   \qquad \qquad \qquad  C_{h,e} = \frac{q_w}{\rho_{\infty} c_p U_{\infty} \left( T_{0,e} - T_w \right)}
\label{eqn:Ch_aw_e}
\end{equation}
where $T_r$ is the recovery temperature at the wall and $T_{0,e}$ is the total temperature with respect to the boundary layer edge conditions. For higher $T_w/T_r$ approaching 1, both the numerator ($q_w$) and the denominator ($T_r - T_w$) approaches zero. For such cases, $C_{h,e}$ provides a more appropriate form to quantify the heat transfer at the wall. Hence, among the flow cases given in Table~\ref{tab:flow_cases}, $C_{h,e}$ will be used to present results for M6Tw076 case. Note that, due to the presence of a leading--edge shock wave, the effective freestream conditions beneath the shock wave are used to calculate $C_h$ and $C_f$.

\subsection{Single Training Case}
\label{sec:single_training}
\begin{figure} [hbt!]
\centering
\subfloat[Velocity profile]{\includegraphics[width=0.32\textwidth]{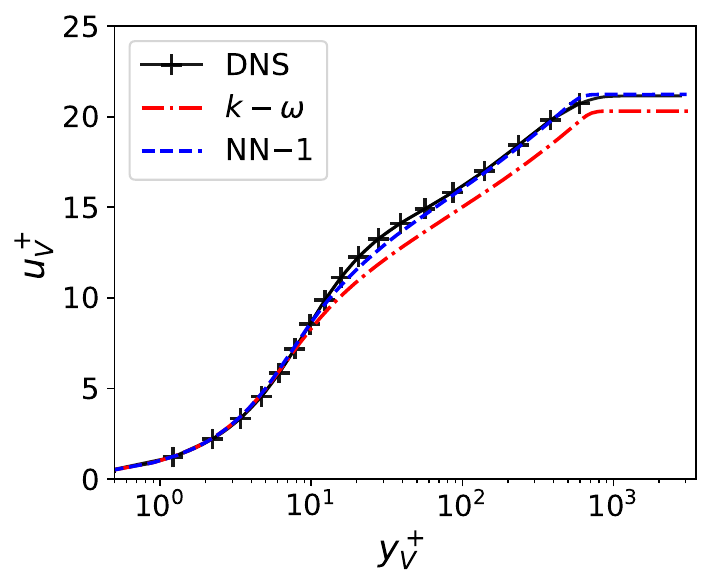}} \qquad 
\subfloat[Temperature profile]{\includegraphics[width=0.32\textwidth]{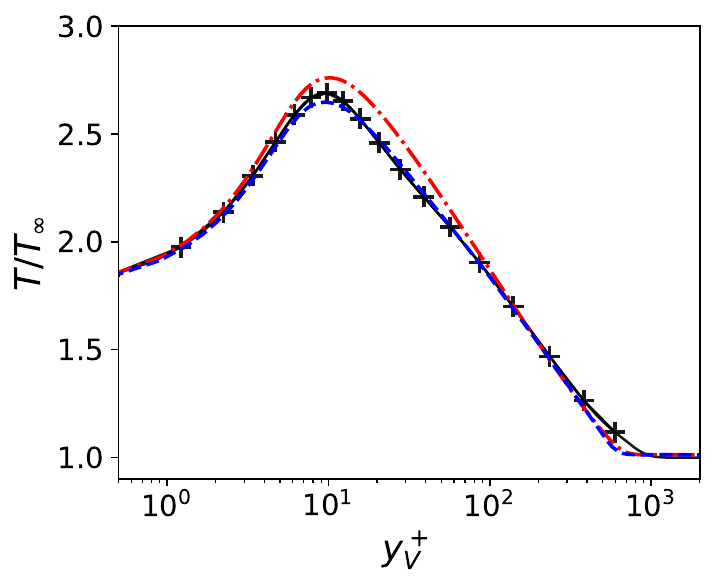}}
\caption{Training results for M6Tw025 flow case, with sparse observation data points ($+$) from DNS are shown.}
\label{fig:single_M6Tw025_uT}
\end{figure}

In this training scenario, the NN--based turbulence model is trained using DNS data for the M6Tw025 flow case and subsequently tested for the remaining four flow cases. Note that the training data corresponds to a single flow case in which the dimensionless wall-temperature input feature is uniform along the flat plate, due to the zero-pressure-gradient flow and fixed wall temperature. Consequently, the input feature of dimensionless wall--temperature assumes a uniform value and is omitted for the single training case, and the remaining six input features are used. The training process of the NN-based turbulence model generates an ensemble of model parameters, with the final model represented by their average. Consequently, the results presented below were obtained using the turbulence model defined by the mean of the ensemble parameters.
The results of this training case are presented in Figure~\ref{fig:single_M6Tw025_uT}, which compares the simulation results obtained using the trained NN--based turbulence model with those from the baseline turbulence model ($k-\omega$). Additionally, the DNS data points used for training are displayed. The results demonstrate that the trained turbulence model predicts velocity and temperature profiles for the training flow case with improved accuracy.

\begin{figure} [hbt!]
\centering
\subfloat[Wall heat transfer coefficients]{\includegraphics[width=0.33\textwidth]{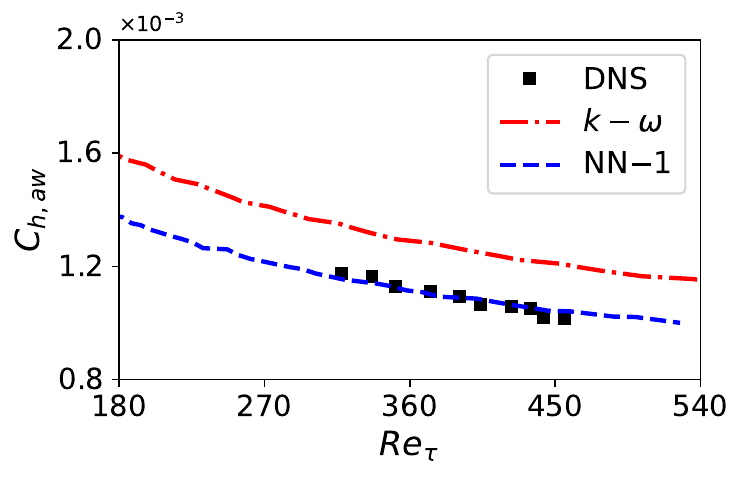} } \qquad 
\subfloat[Skin friction coefficient]{\includegraphics[width=0.33\textwidth]{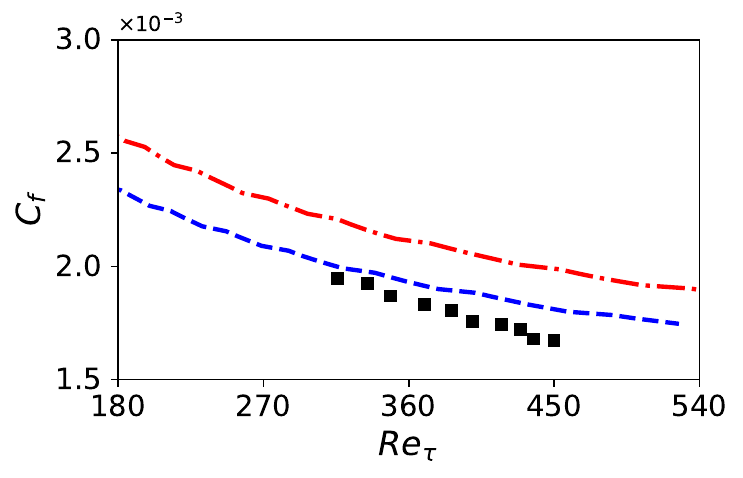}
}
\caption{Quantities of interest at the wall for the training case M6Tw025.}
\label{fig:single_M6Tw025_C}
\end{figure}

\begin{figure} [hbt!]
\centering
\subfloat[Turbulent Prandtl number]{\includegraphics[width=0.31\textwidth]{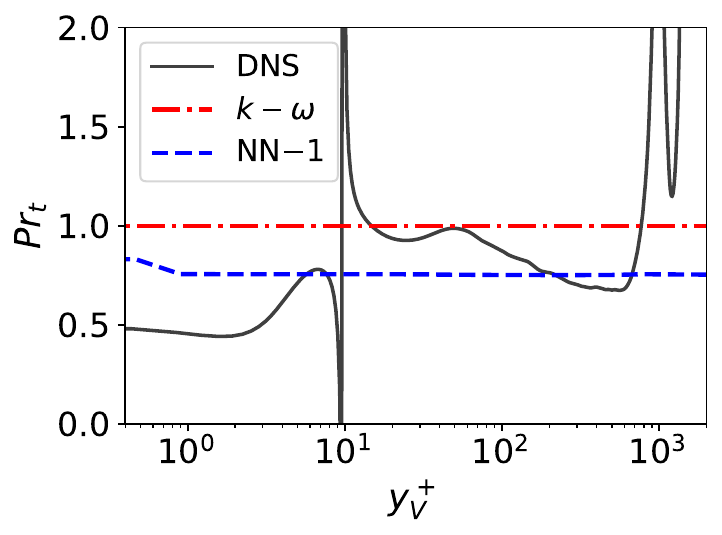}} \qquad 
\subfloat[Scalar coefficient of the first tensor basis]{\includegraphics[width=0.33\textwidth]{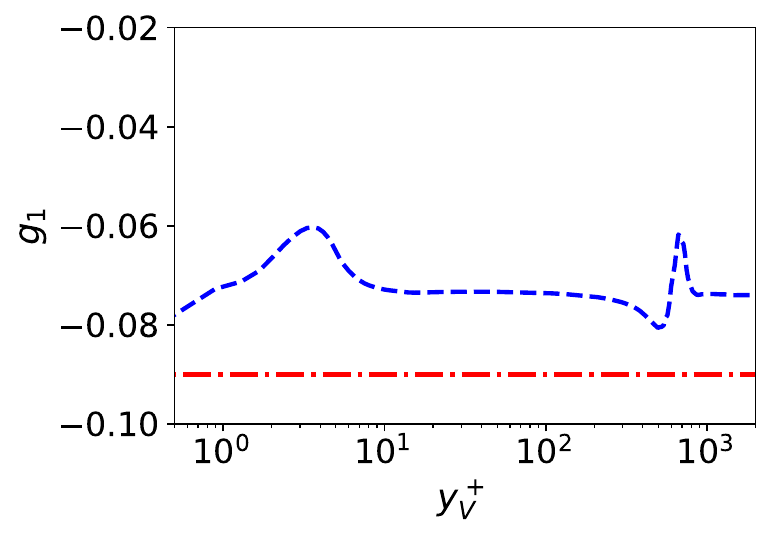}
}
\caption{Variation of learned closure variables in the boundary layer at the sampling location for the training case M6Tw025.}
\label{fig:single_M6Tw025_output}
\end{figure}

The wall heat transfer and skin friction coefficients, shown in Figure~\ref{fig:single_M6Tw025_C}, are also predicted with significantly improved accuracy compared to the baseline model. This improvement can be attributed to the learned turbulence model that predicts the output quantities of $g^{(1)}$ and $Pr_t$ considerably  different from the baseline model, as shown in Figure~\ref{fig:single_M6Tw025_output}. The predicted $g^{(1)}$ exhibits variations with respect to the input features  normal to the wall within the boundary layer region, deviating from the constant value of $-0.09$ assumed by the baseline model. On the other hand, the predicted $Pr_t$ remains constant, albeit different from the baseline model. When compared to the variable $Pr_t$ from the DNS data, the predicted constant value of $Pr_t$ appears to be roughly equal to the mean of the $Pr_t$ profile from DNS data. 
The singular behavior in the $Pr_t$ profile from DNS  at $y_V^+\approx10$ (Fig.~\ref{fig:single_M6Tw025_output}(a)) corresponds to the inflection in the temperature profile at $y_V^+\approx10$ (Fig.~\ref{fig:single_M6Tw025_uT}(b). The singular behavior arises at the inflection point of the temperature profile because both the wall--normal temperature gradient and the turbulent heat flux, terms in the numerator and denominator of the $Pr_t$ definition for DNS data, approach zero.

\begin{figure} [hbt!]
\centering
\subfloat[Initial $Pr_{t} = 0.5$ (constant value)]{\includegraphics[width=0.33\textwidth]{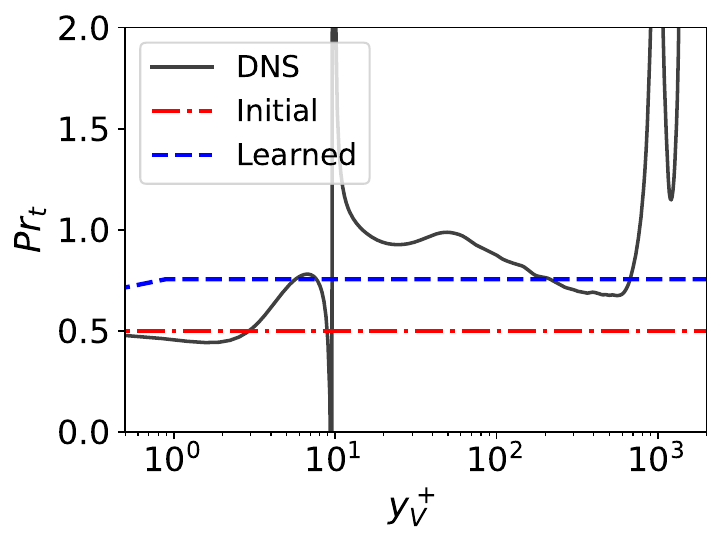}} \qquad 
\subfloat[Initial $Pr_{t}$ varies periodically around 1.0]{\includegraphics[width=0.33\textwidth]{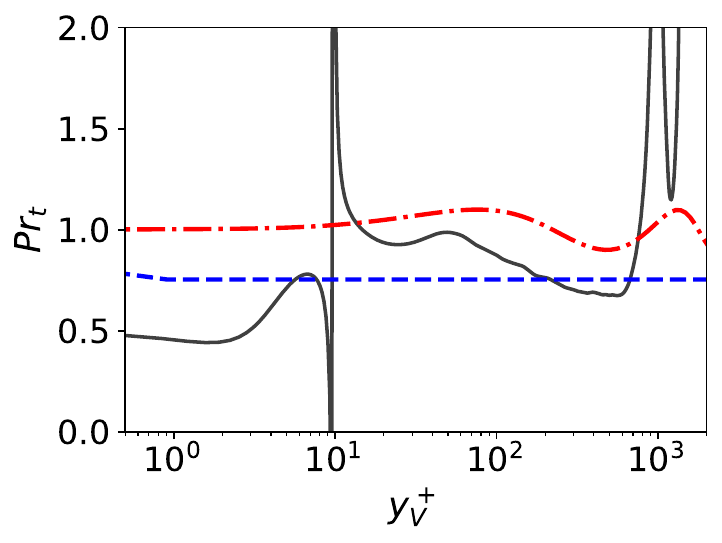}}
\caption{Results of two different training cases with different initial values of turbulent Prandtl number ($Pr_t$).
}
\label{fig:initial_Prt}
\end{figure}

The training of the NN--based turbulence model is also explored by varying the initial values of $Pr_t$ from the previously used value of 1.0. This is accomplished by employing different values of $Pr_t$ during the pre-training of the NN parameters and in the RANS simulation with the baseline model to extract input features for the first training iteration. Two distinct initial values of $Pr_t$ are examined: a uniform value of 0.5 and sinusoidal variation of $Pr_t$ in the wall--normal direction with a mean around 1.0. The results are shown in Figure~\ref{fig:initial_Prt}, where the initial and learned values of $Pr_t$ are compared. In both cases, the predicted $Pr_t$ by the trained turbulence model ultimately converges to the same narrow range of $Pr_t$ values, regardless of the initial values employed for training. 

\begin{figure} [hbt!]
\centering
\subfloat[M6Tw076]{\includegraphics[width=0.31\textwidth]{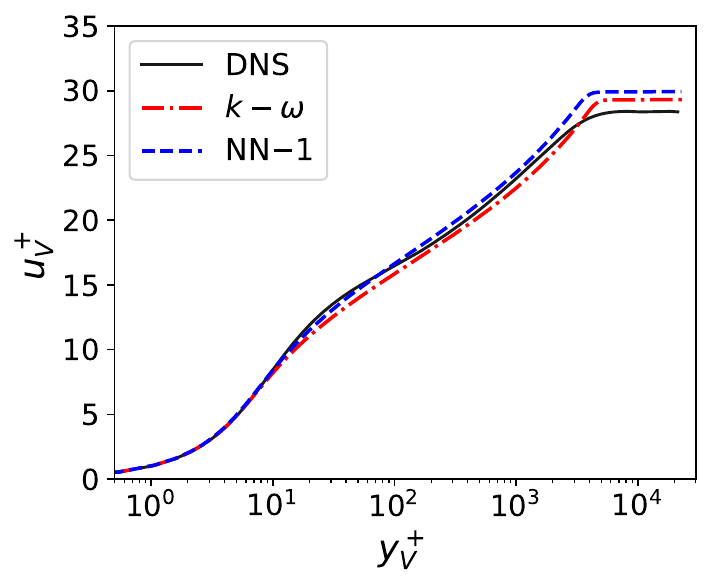}} \qquad
\subfloat[M8Tw048]{\includegraphics[width=0.31\textwidth]{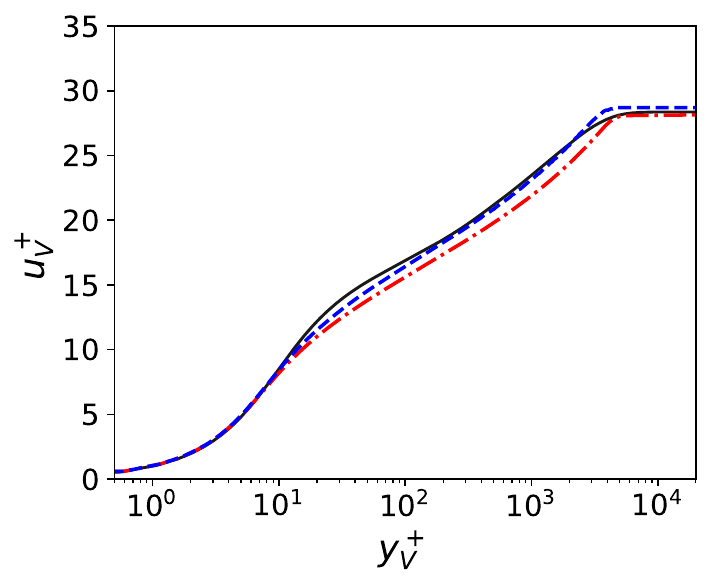}} \\
\subfloat[M11Tw020]{\includegraphics[width=0.31\textwidth]{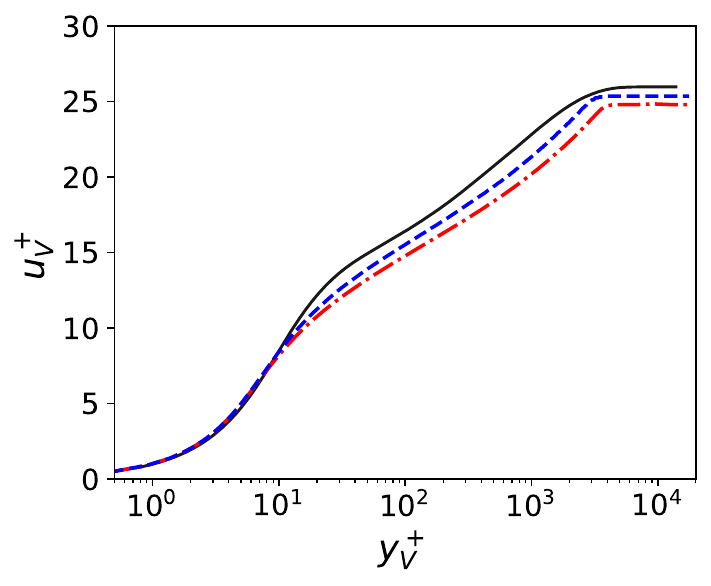}} \qquad
\subfloat[M14Tw018]{\includegraphics[width=0.31\textwidth]{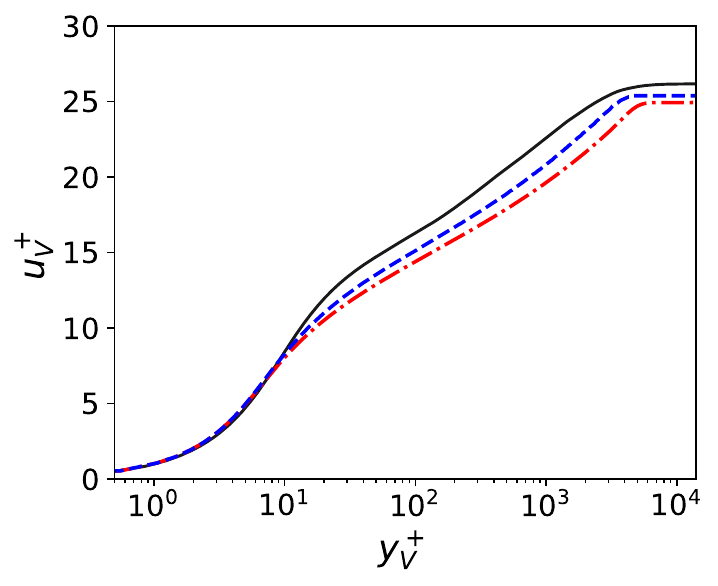}}
\caption{Velocity profiles for the test cases, computed using the NN model trained with M6Tw025 case.}
\label{fig:single_test_uPlus}
\end{figure}

\begin{figure} [hbt!]
\centering
\subfloat[M6Tw076]{\includegraphics[width=0.31\textwidth]{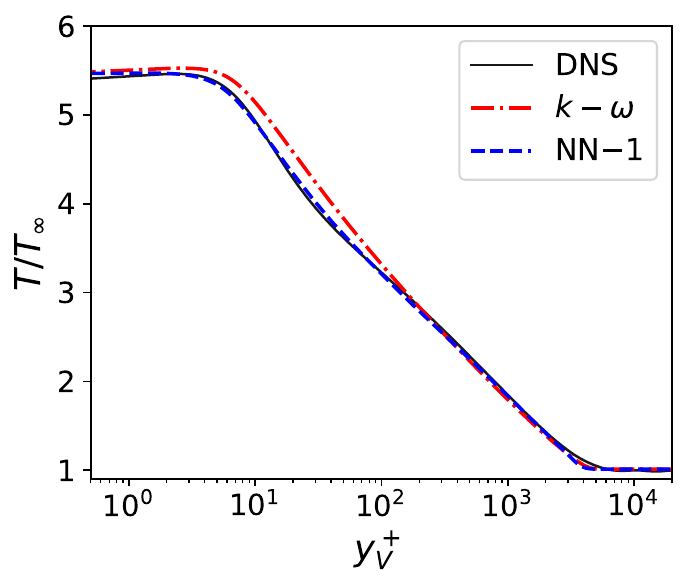}} \qquad
\subfloat[M8Tw048]{\includegraphics[width=0.31\textwidth]{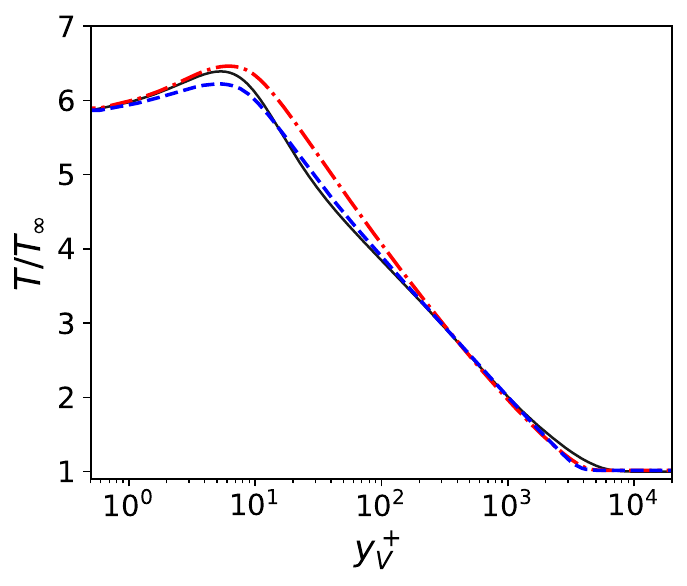}} \\
\subfloat[M11Tw020]{\includegraphics[width=0.31\textwidth]{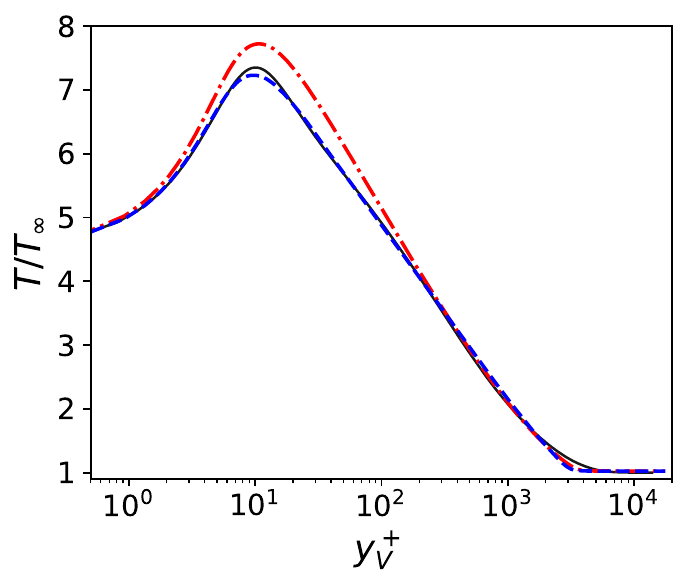}} \qquad
\subfloat[M14Tw018]{\includegraphics[width=0.32\textwidth]{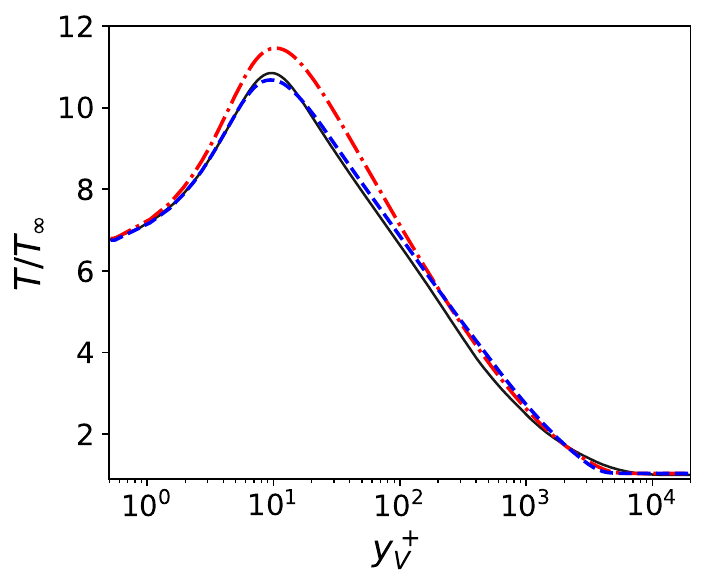}}
\caption{Temperature profiles for the test cases, computed using the NN model trained with M6Tw025 case.}
\label{fig:single_test_T}
\end{figure}

\begin{figure} [hbt!]
\centering
\subfloat[M6Tw076]{\includegraphics[width=0.325\textwidth]{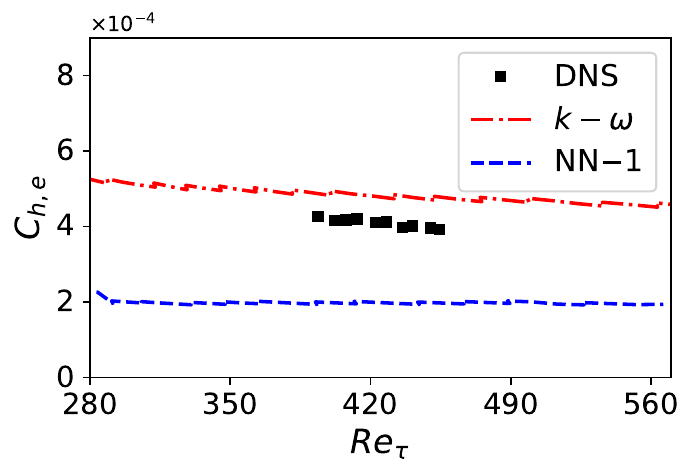}} \qquad
\subfloat[M8Tw048]{\includegraphics[width=0.315\textwidth]{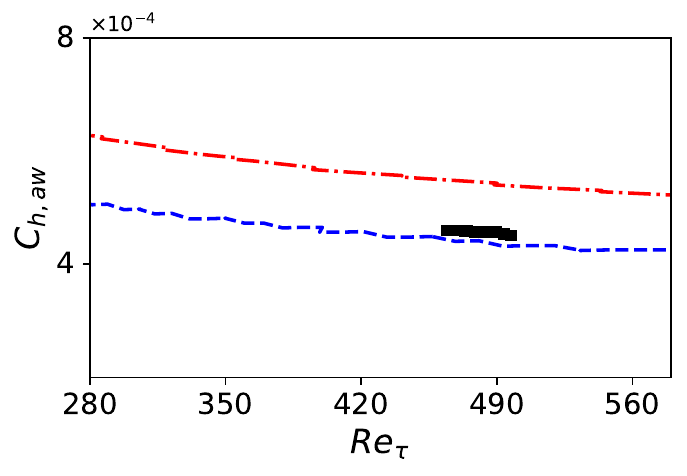}} \\
\subfloat[M11Tw020]{\includegraphics[width=0.325\textwidth]{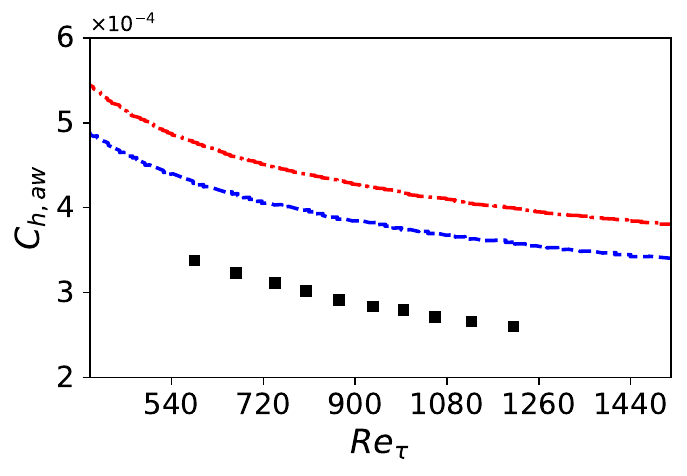}} \qquad
\subfloat[M14Tw018]{\includegraphics[width=0.325\textwidth]{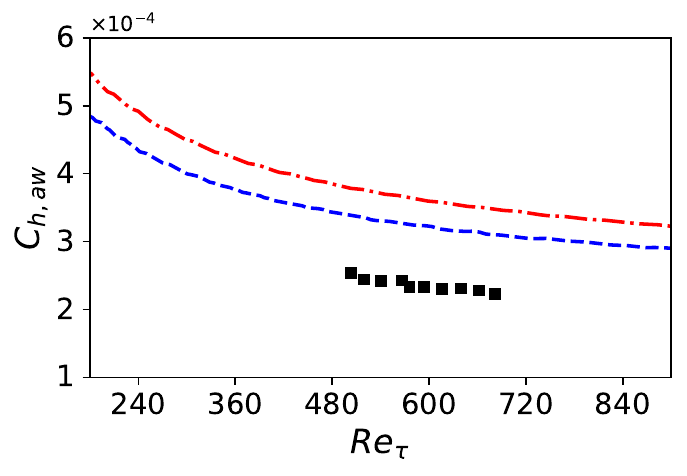}}
\caption{Heat transfer coefficient at the wall for the test cases, computed using the NN model trained with M6Tw025 case.}
\label{fig:single_test_Ch}
\end{figure}

\begin{figure} [hbt!]
\centering
\subfloat[M6Tw076]{\includegraphics[width=0.325\textwidth]{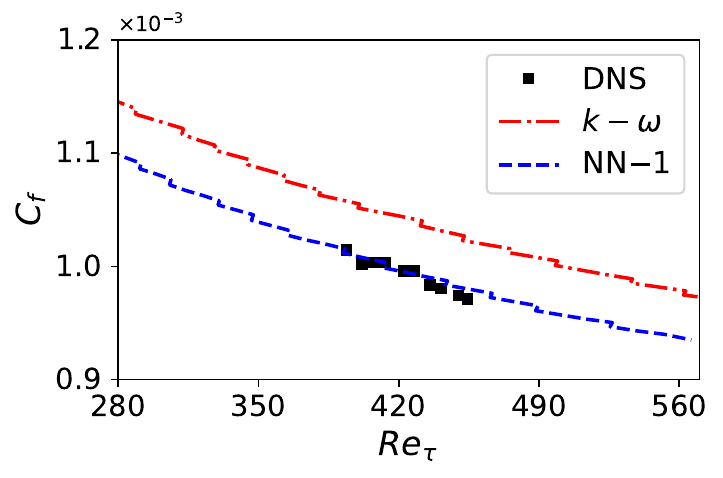}} \qquad
\subfloat[M8Tw048]{\includegraphics[width=0.32\textwidth]{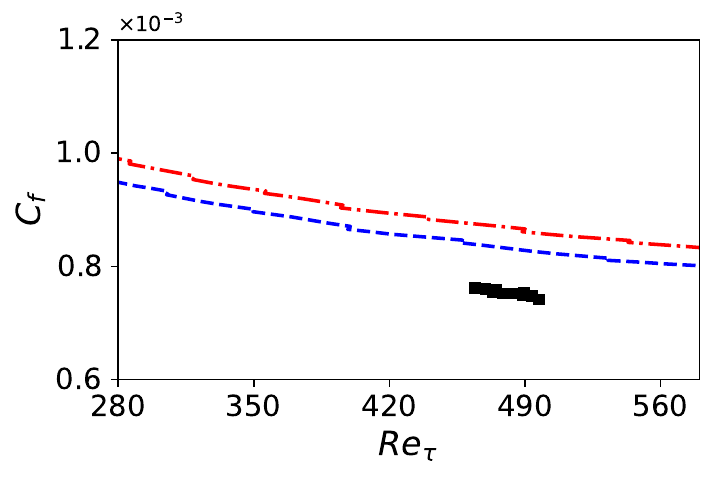}} \\
\subfloat[M11Tw020]{\includegraphics[width=0.33\textwidth]{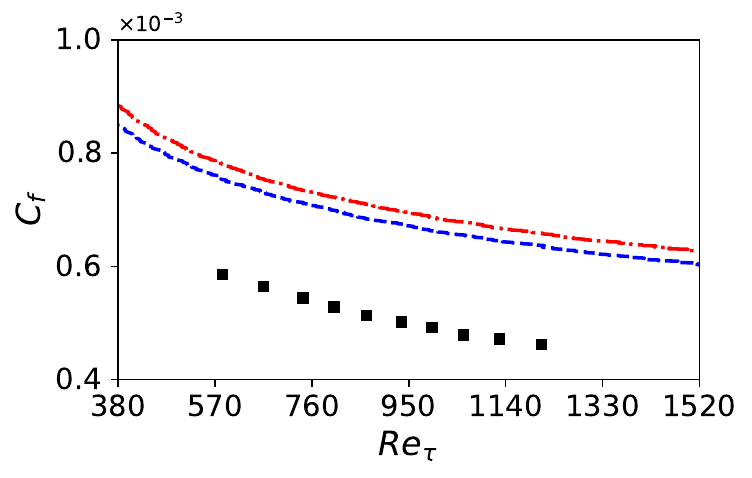}} \qquad
\subfloat[M14Tw018]{\includegraphics[width=0.32\textwidth]{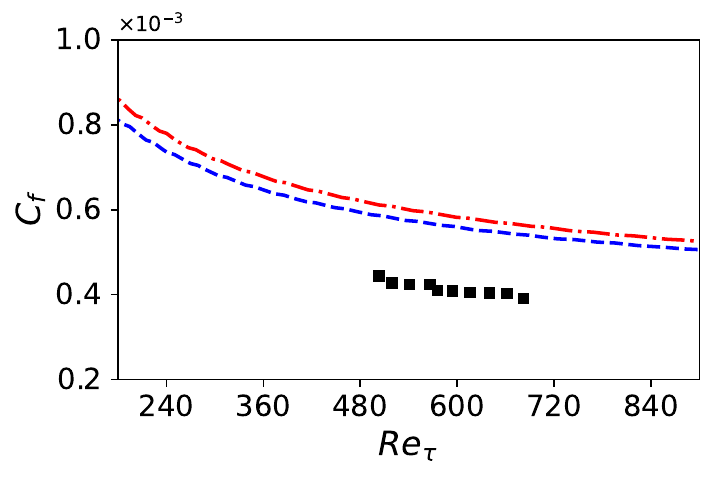}}
\caption{Skin friction coefficient for the test cases, computed using the NN model trained with M6Tw025 case.}
\label{fig:single_test_Cf}
\end{figure}

The turbulence model, trained using DNS data of the M6Tw025 flow case, is tested on the remaining flow cases. Figure~\ref{fig:single_test_uPlus} presents the predicted velocity profiles for these test cases, showing improvements for all cases except for M6Tw076, for which the predicted profile near the BL edge is slightly worse than that of the baseline model. Examining the temperature profiles in Figure~\ref{fig:single_test_T}, a noticeable improvement can be observed, particularly in the peak temperatures within the boundary layer. Wall heat transfer coefficients are shown in Figure~\ref{fig:single_test_Ch}, where improvements can be observed for the M8Tw048, M11Tw020 and M14Tw018 flow cases, whereas it is under--predicted for the M6Tw076 test case. Conversely, for the results of skin--friction coefficient shown in Figure~\ref{fig:single_test_Cf}, only slight improvement is observed for these test cases.

\subsection{Joint Training Case}
\label{sec:joint_training}
Here, we present the results for the joint training case (NN--2), where DNS data from two flow cases are simultaneously used to train the turbulence model. In the single training case (NN--1) presented above, predictions for the test flow cases show sporadic improvements. In particular, the prediction of wall heat transfer and skin friction coefficients for the test flow cases shows only slight improvement, or, in the case of M6Tw076, the wall heat transfer coefficient is significantly underpredicted. Therefore, a joint training case is considered here, where DNS data from two flow cases, namely M6Tw076 and M14Tw018, are used for training. The choice of these two flow cases for training is made in order to provide observational data from two extremes of Mach numbers, as well as wall--to--recovery temperature ratios among the available list of flow cases. A similar number of observation points from both flow cases, with a similar distribution throughout the BL, are used for joint training.

\begin{figure} [hbt!]
\centering
\subfloat[M6Tw076]{\includegraphics[width=0.33\textwidth]{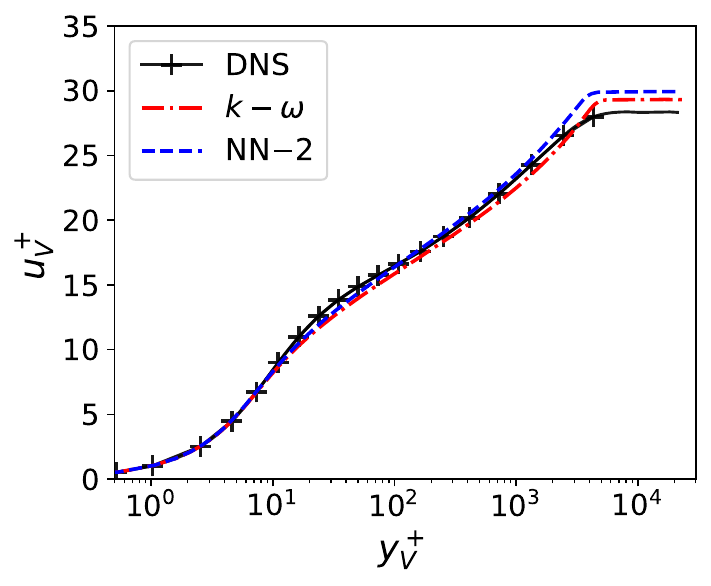} \qquad 
\includegraphics[width=0.32\textwidth]{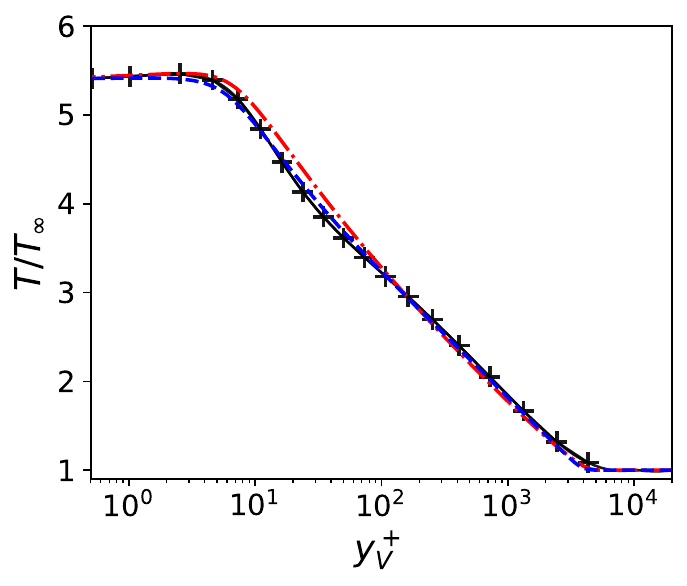}} \\
\subfloat[M14Tw018]{\includegraphics[width=0.33\textwidth]{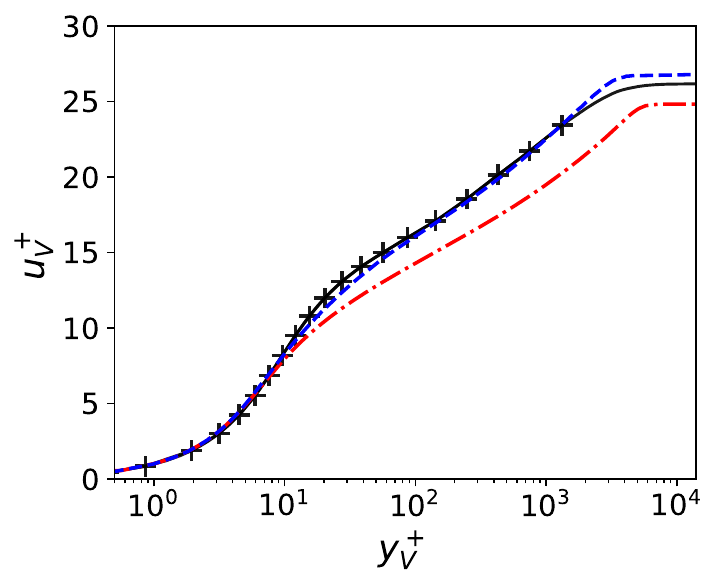} \qquad 
\includegraphics[width=0.325\textwidth]{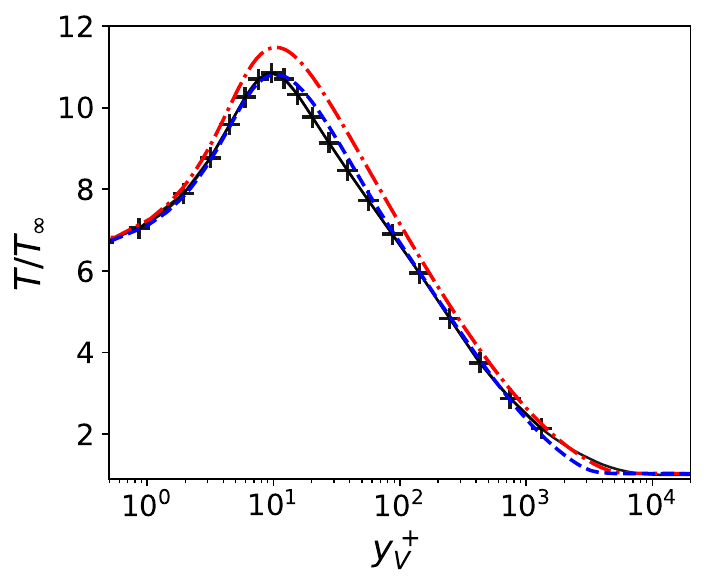}} 
\caption{Joint training results where DNS data of two flow cases is simultaneously used for training. DNS data points (+) of velocity and temperature used for training for both cases are shown as well.}
\label{fig:joint_training}
\end{figure}

Results of the two training flow cases are shown in Figure~\ref{fig:joint_training}, where the velocity and temperature profiles computed using the trained turbulence model are plotted against those obtained with the baseline turbulence model ($k$-$\omega$). Training data points from DNS of the respective flow cases are also displayed. For the M14Tw018 flow case, the trained turbulence model is able to predict velocity and temperature profiles with significantly improved accuracy. For the M6Tw076 flow case, where the baseline model already predicts the flow reasonably well, the predictions of the velocity and temperature profiles showed little to no improvement.
Since the baseline model can reasonably predict the velocity profile of the M6Tw076 flow case, it can be deduced that the standard value of $g_1=-0.09$ performs well for $T_w/T_r$ relatively closer to 1. As the $T_w/T_r$ decreases, such as in the case of M14Tw018, a better agreement of the velocity profile with the DNS data requires a lower value of $g_1$ in terms of absolute number (i.e. $|-0.05| < |-0.09|$).

\begin{figure} [hbt!]
\centering
\includegraphics[width=0.36\textwidth]{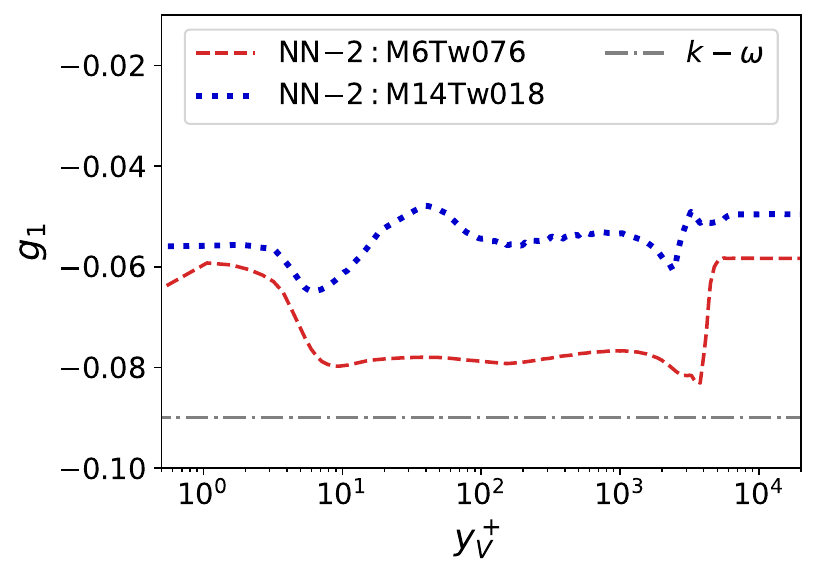} \qquad 
\includegraphics[width=0.34\textwidth]{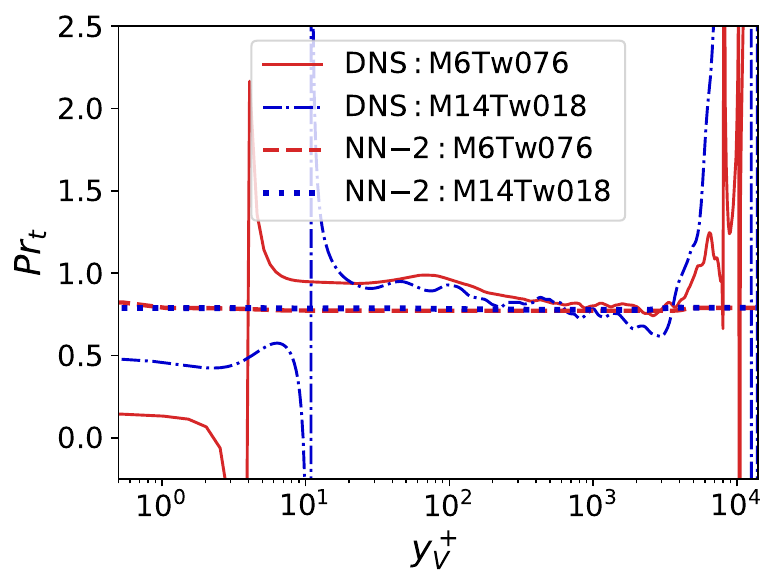}
\caption{Predicted output quantities ($Pr_t$ and $g_1$) for M6Tw076 and M14Tw018 flow cases at their respective sampling locations.}
\label{fig:analysis_outputs}
\end{figure}

This observation is further confirmed by the learned values of $g_1$, which differ significantly between M6Tw076 and M14Tw018, as shown in Fig.~\ref{fig:analysis_outputs}. The model clearly distinguishes between the two flow cases, with the learned values of $g_1$ in the boundary layer being mostly falling within a narrow range around $-0.08$ for M6Tw076 and closer to $-0.055$ for M14Tw018. Note that the values of $g_1$ in the viscous sublayer are not critical, as the turbulent kinetic energy ($k$) is zero, yielding zero values for $\mu_t$ in the viscous sublayer. (Eq.~\ref{eqn:nut}).
Additionally, a test flow case M5Tw091 with higher $T_w/T_r$ of 0.91 is also considered. Results for this test flow case, as presented in Appendix A, further emphasize that as $T_w/T_r$ gets closer to 1, a $g_1$ value around -0.09 is observed as more appropriate for better predictions of mean flow quantities. Conversely, the learned values of $Pr_t$ are predicted to be the same for the two training flow cases and show no variation in the wall--normal direction, as shown in Fig.~\ref{fig:analysis_outputs}. This is in contrast to DNS data which indicates a difference in $Pr_t$ profiles for two flow cases, particularly near the wall. 

\begin{figure} [hbt!]
\centering
\subfloat[M6Tw025]{\includegraphics[width=0.31\textwidth]{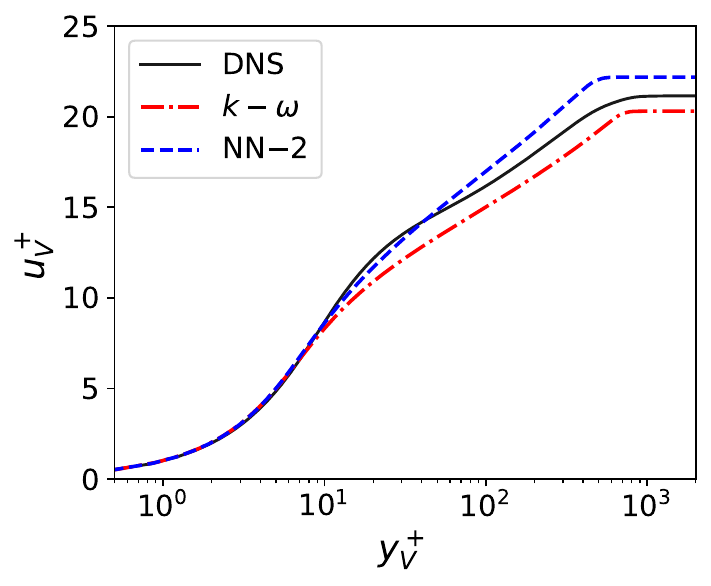}} \ \ 
\subfloat[M8Tw048]{\includegraphics[width=0.31\textwidth]{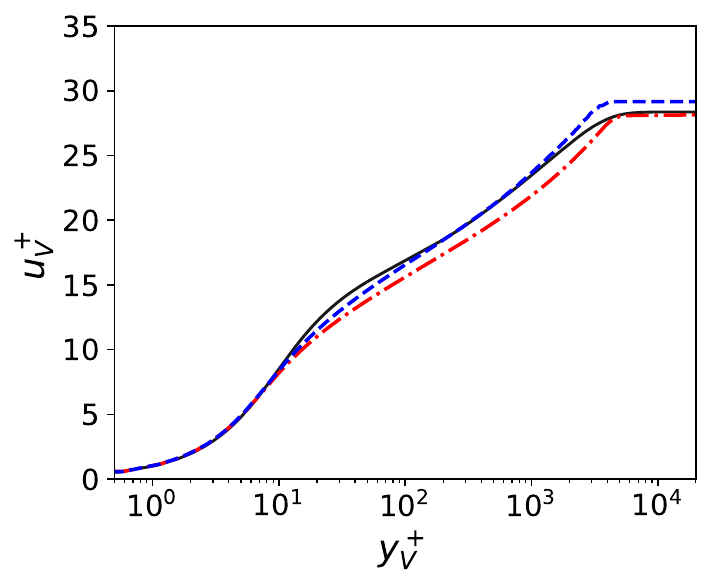}} \ \ 
\subfloat[M11Tw020]{\includegraphics[width=0.31\textwidth]{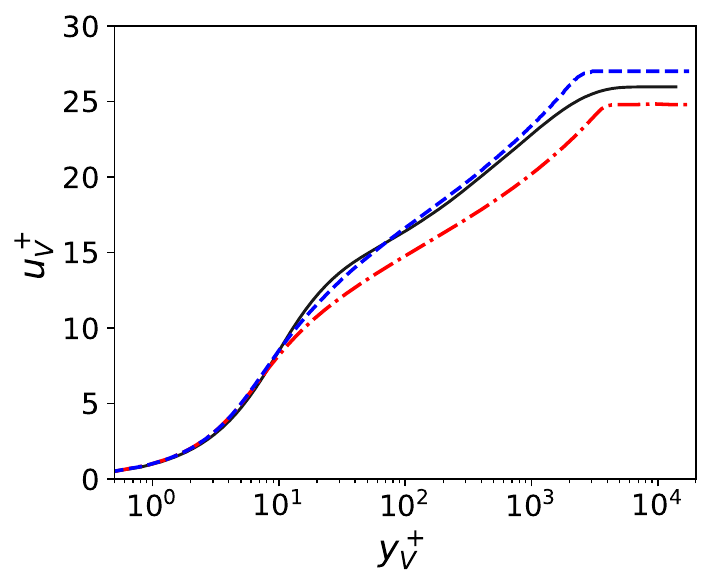}}
\caption{Velocity profiles for the test cases, with NN model jointly trained with M6Tw076 and M14Tw018 data}
\label{fig:joint_test_uPlus}
\end{figure}

\begin{figure} [hbt!]
\centering
\subfloat[M6Tw025]{\includegraphics[width=0.315\textwidth]{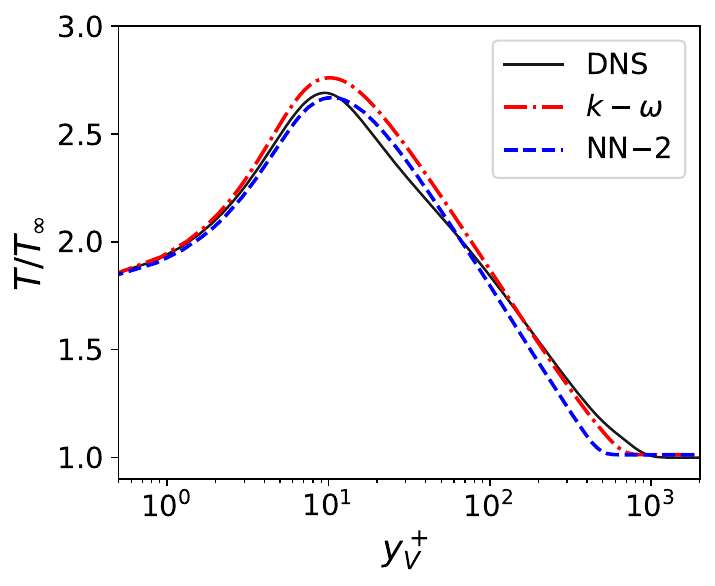}} \ \ 
\subfloat[M8Tw048]{\includegraphics[width=0.3\textwidth]{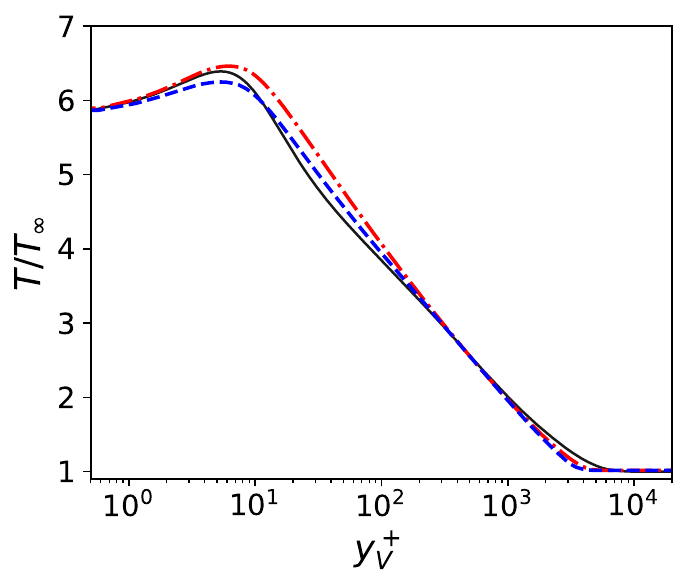}} \ \ 
\subfloat[M11Tw020]{\includegraphics[width=0.3\textwidth]{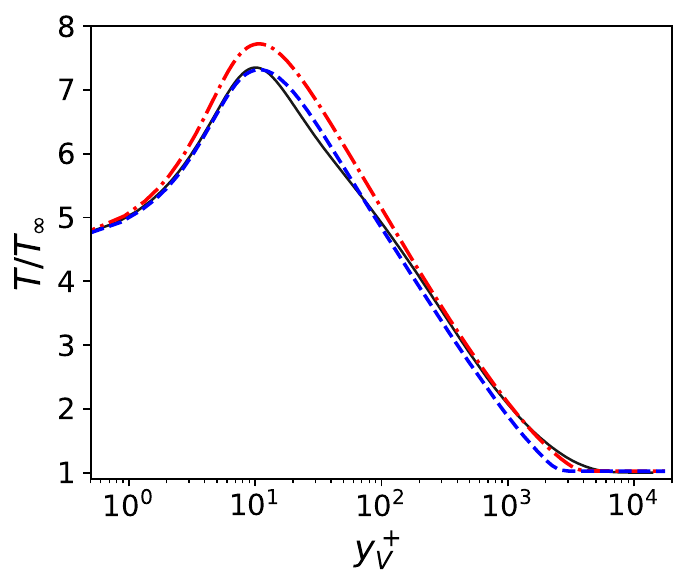}}
\caption{Temperature profiles for the test cases, with NN model jointly trained with M6Tw076 and M14Tw018 data}
\label{fig:joint_test_T}
\end{figure}

The jointly trained turbulence model is tested on the remaining three flow cases. For the velocity profiles shown in Fig.\ref{fig:joint_test_uPlus}, some improvement is observed for M11Tw020, whereas the prediction for M6Tw025 is worse compared to the single training case. The predictions for M6Tw025 are analyzed in more detail in Sec.\ref{sec:results}C. For the temperature profiles shown in Fig.~\ref{fig:joint_test_T}, no significant improvement is observed compared to the single training case.

\begin{figure} [hbt!]
\centering
\subfloat[M6Tw025]{\includegraphics[width=0.33\textwidth]{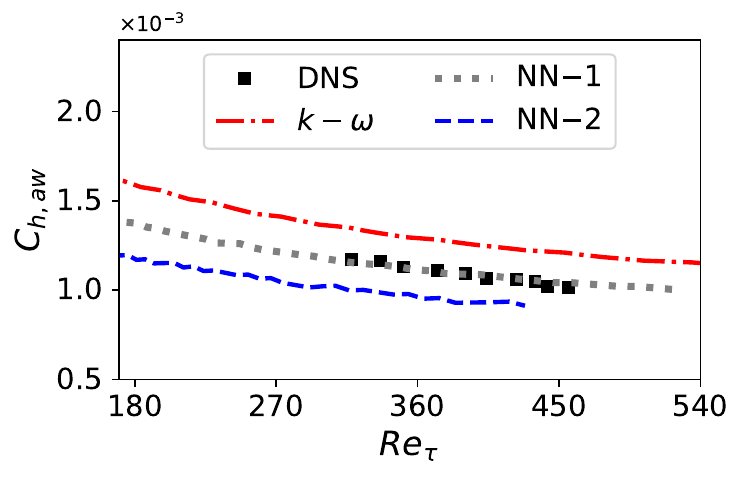}} \
\subfloat[M6Tw076]{\includegraphics[width=0.315\textwidth]{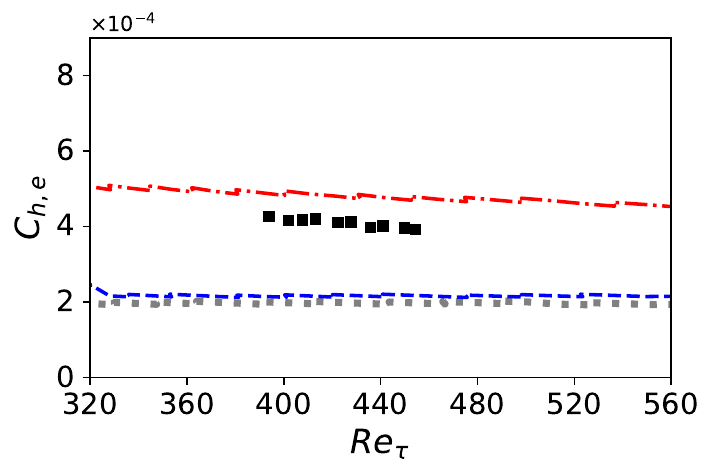}} \
\subfloat[M8Tw048]{\includegraphics[width=0.305\textwidth]{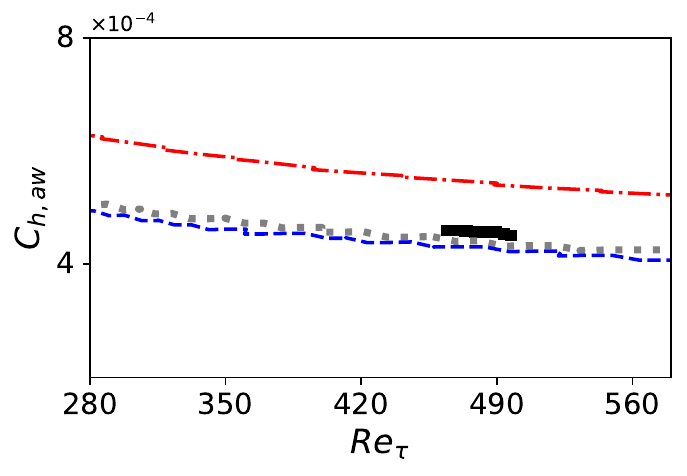}} \
\subfloat[M11Tw020]{\includegraphics[width=0.31\textwidth]{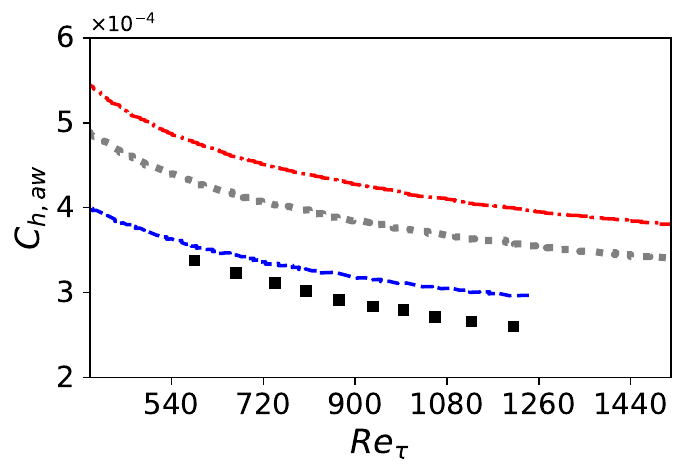}} \ 
\subfloat[M14Tw018]{\includegraphics[width=0.31\textwidth]{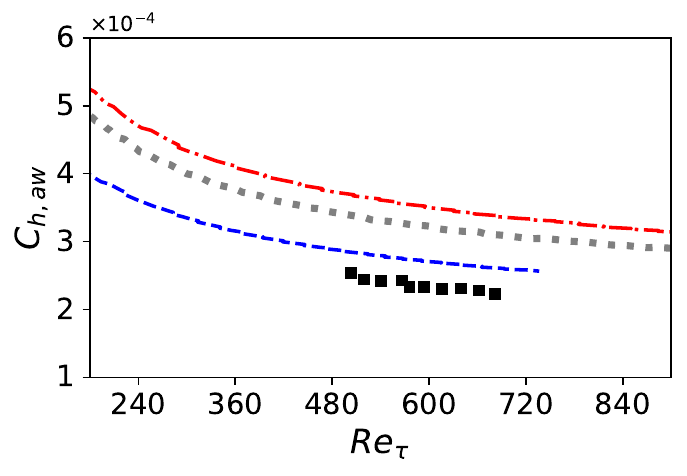}}
\caption{Heat transfer coefficient at the wall for the test cases, with NN model jointly trained with M6Tw076 and M14Tw018 data}
\label{fig:joint_Ch}
\end{figure}

\begin{figure} [hbt!]
\centering
\subfloat[M6Tw025]{\includegraphics[width=0.325\textwidth]{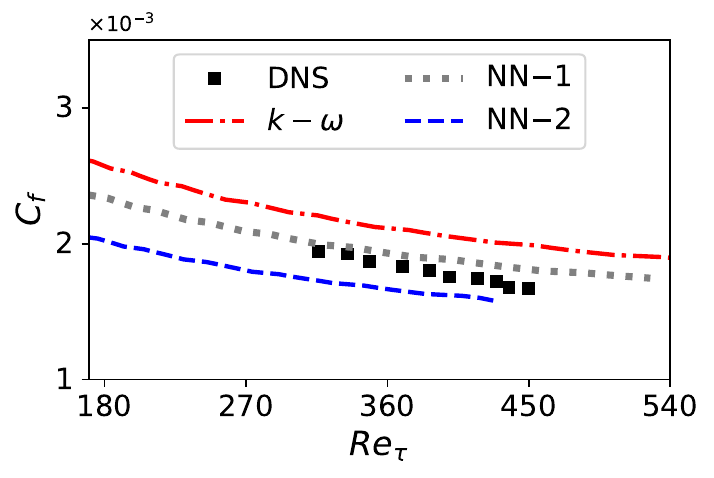}} \
\subfloat[M6Tw076]{\includegraphics[width=0.325\textwidth]{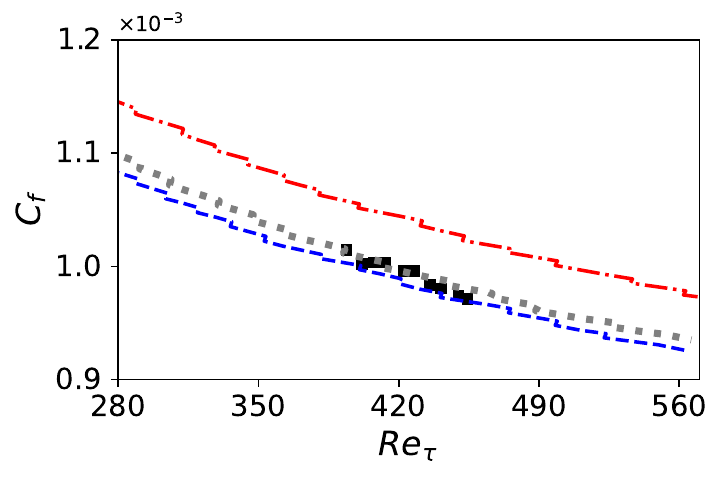}} \
\subfloat[M8Tw048]{\includegraphics[width=0.32\textwidth]{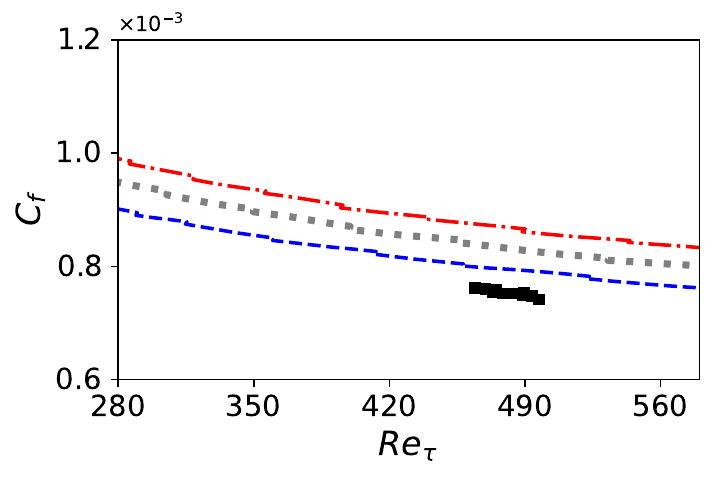}} \
\subfloat[M11Tw020]{\includegraphics[width=0.34\textwidth]{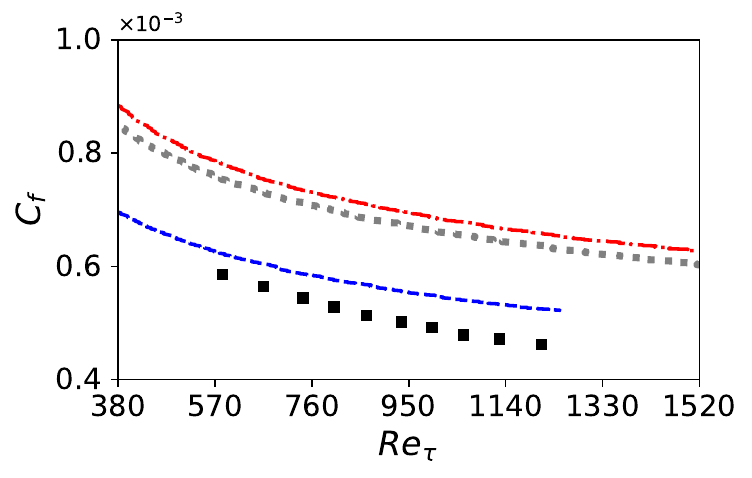}} \ 
\subfloat[M14Tw018]{\includegraphics[width=0.33\textwidth]{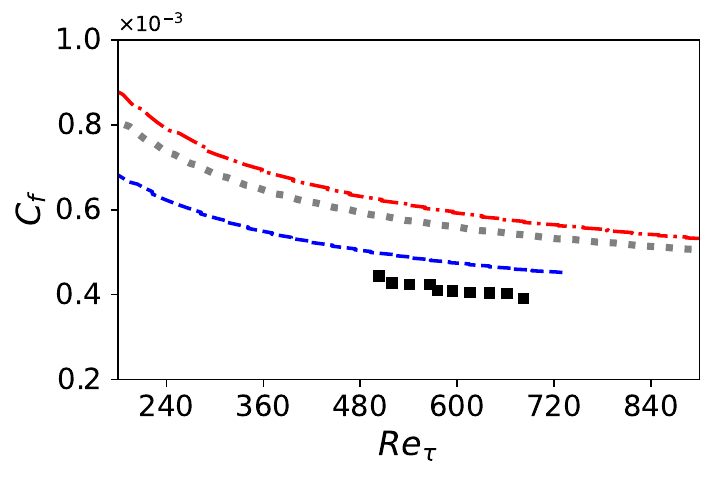}}
\caption{Skin friction coefficient for the test cases, with NN model jointly trained with M6Tw076 and M14Tw018 data}
\label{fig:joint_Cf}
\end{figure}

Results for the wall heat transfer and skin friction coefficients are presented in Figs.~\ref{fig:joint_Ch} and \ref{fig:joint_Cf}. A comparison is made between two trained neural network (NN) models: the single-training-case model, NN–1, trained using data from one flow case (M6Tw025), and the joint-training-case model, NN–2, trained using data from two flow cases (M6Tw076 and M14Tw018). Overall, the joint training case demonstrates better predictions for wall quantities. Compared to NN–1, the predictions for the wall heat transfer coefficient ($C_h$) in Fig.\ref{fig:joint_Ch} significantly improve with NN–2 for the flow cases M11Tw020 and M14Tw018, show no significant improvement for M6Tw076, and worsen for M6Tw025. Similarly, the predictions for the skin friction coefficient ($C_f$) in Fig.\ref{fig:joint_Cf} significantly improve with NN–2 for M8Tw048, M11Tw020, and M14Tw018, show slight changes for M6Tw076, and worsen for M6Tw025.

\subsection{Discussion}
The results presented above demonstrate the ability of learning an NN-based turbulence model using sparse observation data of mean flow quantities. The trained turbulence models resulted in significantly improved accuracy compared to the baseline model when computing mean flow quantities. This can be compared with the recently proposed compressibility correction~\cite{danis2022kOmegaCorr} for cold--wall conditions, where little to no improvement was observed in the computation of the mean temperature field while some improvement was observed for velocity field. However, the generalization of the trained turbulence model to test flow cases with different Mach numbers and wall--to--recovery temperature ratios does not \textit{consistently} lead to improvements in the computation of mean flow quantities. Overall, the prediction of wall quantities of interest ($C_h$ and $C_f$) by the jointly trained turbulence model (NN--2) shows significant improvement compared to the baseline model.

\begin{figure} [hbt!]
\centering
\includegraphics[width=0.37\textwidth]{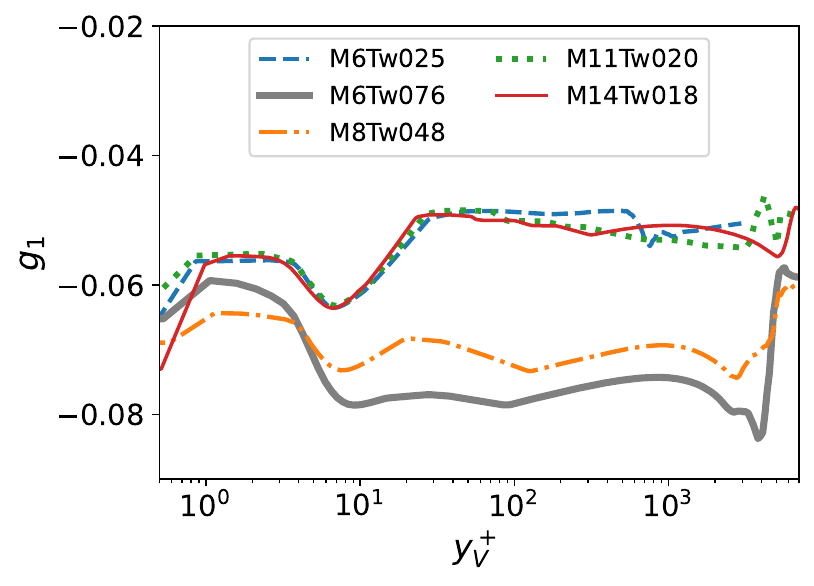}
\caption{Output quantity (scalar invariant $g_1$) as predicted at the respective sampling locations for each flow case by the jointly trained NN--2 model.}
\label{fig:joint_g1}
\end{figure}

Turbulent flow behavior varies significantly under different cold-wall conditions, and incorporating the dimensionless wall temperature parameter into the input feature set greatly enhances the ability to capture this variation. In this regard, the jointly trained model effectively differentiates between cold--wall conditions by adjusting the Reynolds stress term (as $\mu_t$ depends on $g_1$ in Eq.~\ref{eqn:nut}). As shown in Figure~\ref{fig:joint_g1}, a decrease in $T_w/T_r$ leads to smaller absolute values of $g_1$ in both the buffer layer and the log--law region of the boundary layer. In the viscous sublayer, where turbulent fluctuations are heavily suppressed (i.e., $k \approx 0$ in Eq.~\ref{eqn:nut}), $\mu_t$ approaches zero, regardless of $g_1$ values.

\begin{figure} [hbt!]
\centering
\subfloat[Scalar invariant $g_1$]{\includegraphics[width=0.345\textwidth]{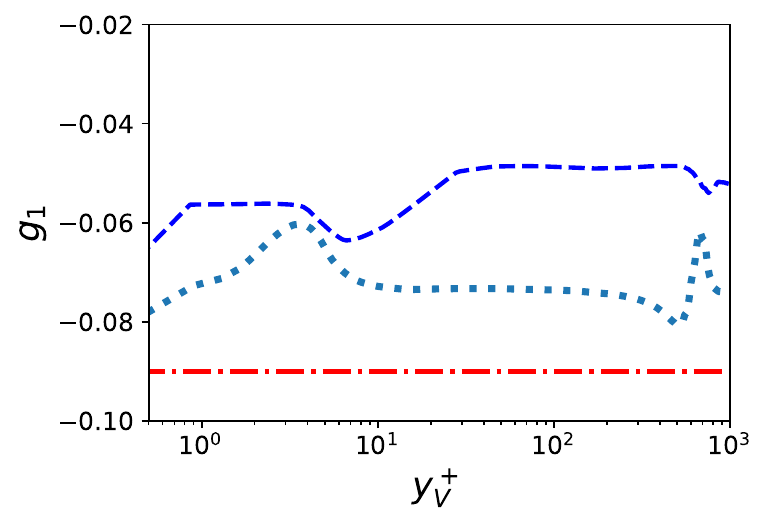}} 
\subfloat[Reynolds shear stress]{\includegraphics[width=0.33\textwidth]{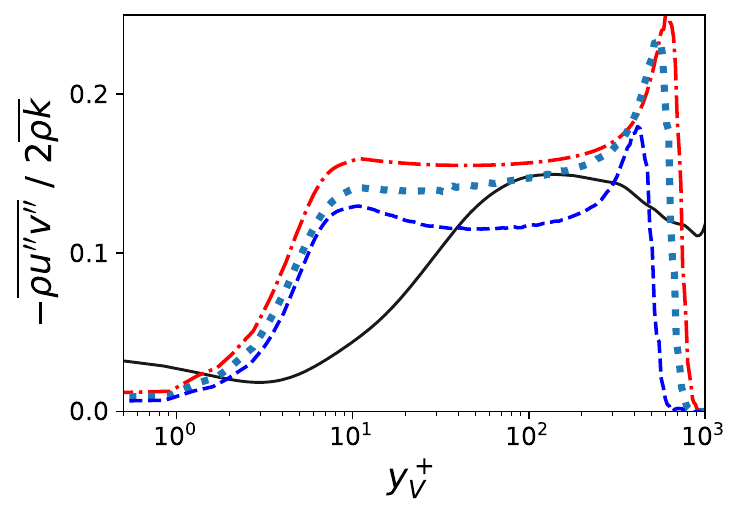}} 
\subfloat[Wall--normal velocity profile]{\includegraphics[width=0.325\textwidth]{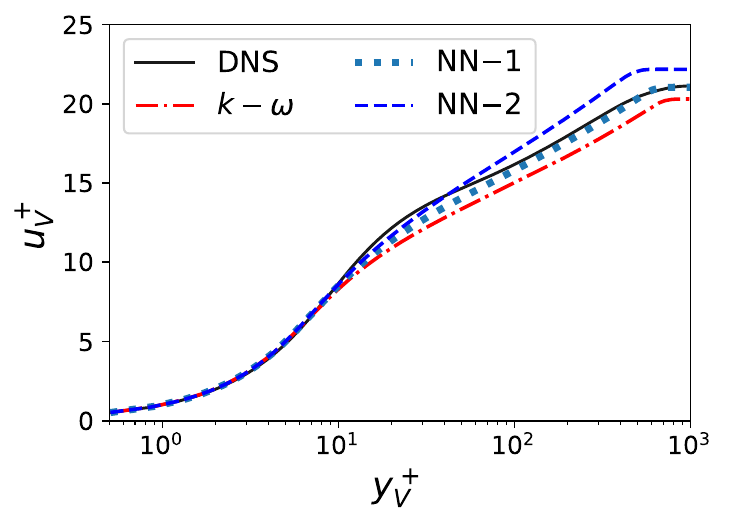}}
\caption{Comparison of the predicted quantities from DNS, the baseline model, and both NN-trained models for the M6Tw025 flow case.}
\label{fig:M6Tw025_analysis}
\end{figure}

To further assess the influence of the predicted $g_1$ values on Reynolds shear stress and the mean velocity profile, we analyze the M6Tw025 flow case. In the single-training-case model (NN--1), which was trained using M6Tw025 data, the velocity profile is predicted with high accuracy, and the absolute values of $g_1$ fall within the range of 0.07 to 0.08. In contrast, the joint-training-case model (NN--2), where M6Tw025 is treated as a test case, predicts a less accurate velocity profile, with $g_1$ values ranging from 0.05 to 0.06. This comparison is illustrated in Fig.~\ref{fig:M6Tw025_analysis}, which also presents the corresponding wall--normal distribution of Reynolds shear stress.

Larger absolute values of $g_1$ correspond to increased eddy viscosity $\mu_t$ and Reynolds shear stress, which in turn represents stronger turbulent mixing and enhanced momentum diffusion. This results in a flatter velocity profile, as momentum is distributed over a wider region; effectively increasing the boundary layer thickness and reducing the need for steep velocity gradients. The baseline model predicts consistently high shear stress, producing the lowest velocity gradient, particularly in the log--law region. In contrast, the NN--2 model underpredicts shear stress, leading to the steepest velocity gradient among the models. The NN--1 model provides the closest match to DNS predictions in the log--law region, but exhibits notable discrepancies in the buffer region.
According to the DNS data, a sharp rise in Reynolds shear stress between $8 < y_V^+ < 80$ results in a gradual decrease in the velocity gradient, creating a visible bump in the velocity profile within this range. Because NN--1 fails to accurately capture the shear stress behavior of sharp rise in this region, it is also unable to accurately replicate the bump in the predicted velocity profile. 

Although the trained models predict $Pr_t$ values that differ from the baseline model, they exhibit limitations in capturing the variability of $Pr_t$ across different $T_w/T_r$ conditions as well as in the wall--normal direction, as illustrated in Fig.\ref{fig:analysis_outputs}. Investigations do not indicate that the neural network architecture or model size is the cause of this limitation. Increasing the network size to enhance expressiveness does not improve the model’s ability to learn the complex wall--normal variation of $Pr_t$. Similar challenges in capturing nonlinear variations of closure variables have also been reported in other work using the ensemble Kalman method\cite{zhang2022jfm}.
An alternative approach is to incorporate both \textit{direct} data of $Pr_t$ and \textit{indirect} data of velocity and temperature during training. Initial investigations indicate that, although this method can capture some degree of the nonlinear variation in $Pr_t$'s wall--normal profile, it substantially compromises the accuracy of the predicted mean flow quantities.

\begin{figure} [hbt!]
\centering
\includegraphics[width=0.41\textwidth]{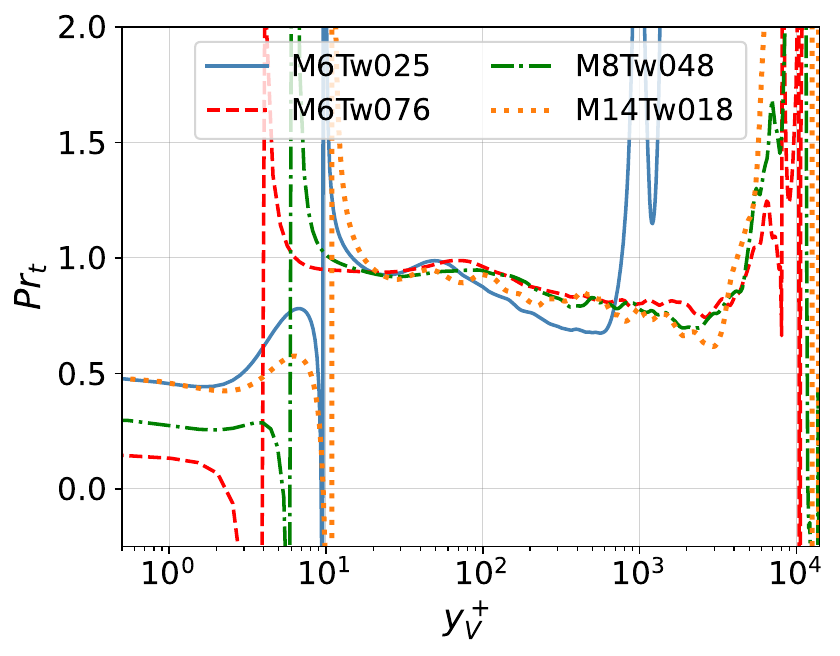}
\caption{Turbulent Prandtl number profiles from DNS database of different flow cases.}
\label{fig:Prt_dns}
\end{figure}

It is also worth pointing out that the $Pr_t$ obtained from DNS data is a derived quantity, based on resolved Reynolds stresses and turbulent heat fluxes $\left( P r_t \equiv\left(\overline{\rho u^{\prime} w^{\prime}}(\partial \bar{T} / \partial y)\right) /\left(\overline{\rho w^{\prime} T^{\prime}}(\partial \bar{u} / \partial y)\right)\right)$. 
In RANS simulations, $Pr_t$ is a modeled quantity, and the functional forms assumed in the modeling of turbulent heat fluxes inherently introduce constraints. From the modeling of the turbulent heat flux term, as expressed in Eq.\ref{eqn:turbheatFlux}, it can be observed that as $\mu_t$ approaches zero near the wall, any learned value of $Pr_t$ will have a negligible impact on the prediction of wall heat flux. However, in the DNS data shown in Fig.\ref{fig:Prt_dns}, the near-wall $Pr_t$ is a function of $T_w/T_r$ and near--wall $Pr_t$ decreases as the ratio $T_w/T_r$ increases.

The analysis of neural network architecture and model size indicates that deficiencies in predicting closure variables are not due to a lack of expressiveness in the model. A set of configurations were tested, combining different numbers of hidden layers (5 or 10) with varying neurons per layer (10, 20, or 40). The results show that a network with 10 hidden layers and 10 neurons per layer offers an optimal balance between model complexity and accuracy in predicting mean flow quantities. Notably, increasing the network size does not improve the ability to capture variations in $Pr_t$ either in the wall--normal direction or across different $T_w/T_r$ conditions.

For flow cases with $T_w/T_r$ approaching 1, the predicted temperature profiles are as accurate as those of the remaining flow cases, particularly near the wall. However, the $C_{h,e}$ predictions for M6Tw076 (Fig.\ref{fig:single_test_Ch}a and \ref{fig:joint_Ch}b) and M5Tw091 (Fig.\ref{fig:joint_test_M5Tw091}c) do not exhibit similar conformity with the DNS data. This discrepancy can be attributed to the relatively small wall heat flux values in such cases, where $C_{h,e}$ shows high sensitivity to small variations in temperature profiles, and numerical errors are likely to be large. This sensitivity is also evident in the Reynolds analogy factor ($R_{af}$) values for the corresponding DNS data, as highlighted in Appendix B. Despite scaling the heat transfer coefficient with respect to boundary layer edge conditions, this discrepancy persists for flow cases with relatively higher $T_w/T_r$ values. Consequently, a different scaling should be formulated, distinct from those presented in Eq.~\ref{eqn:Ch_aw_e}.

The training cost with the ensemble Kalman method is directly related to the computational cost of the RANS simulation for the specific hypersonic flow case. Since the forward propagation of the ensemble is performed in parallel during each training iteration, the training cost can vary significantly with respect to the flow conditions, size of the computational grid, and the resources available for parallel computation. In comparison, the cost associated with the analysis and parameter updates during each iteration is negligible. Since the baseline solution serves as the starting point for training, performing forward propagation for a fraction (approximately $20\%$) of the total simulation time required for the convergence of the baseline solution is sufficient during each iteration. The training costs for both single and joint training cases are summarized in Table~\ref{tab:train_cost}, which indicates the compute resources (in terms of processor cores) and associated wall times required to achieve convergence with the specified number of iterations. Prediction with the trained turbulence model is fairly efficient, as the baseline solution is used as the initial condition, and forward propagation is performed for a similar duration of simulation time as during training in each iteration.

\begin{table}[htbp]
\centering
\caption{Summary of the computational cost for each of the training cases. \label{tab:train_cost}}
\begin{tabular}{l c c c c} \hline
Training Case & Samples & Cores per sample & Iterations & Wall time \\
\hline
Single~(\S \ref{sec:single_training}) & 20 & 4 & 35  & 66 hrs. \\ 
Joint~(\S \ref{sec:joint_training}) & 40 & 8 & 18 & 139 hrs. \\ \hline
\end{tabular}
\end{table}

\section{Conclusion}
\label{sec:conclusion}
This work presents data--driven turbulence modeling for zero-pressure gradient hypersonic flows with cold--wall conditions. The turbulence model is learned to provide closure for Reynolds stress and a variable turbulent Prandtl number, with an emphasis on improving the prediction of near--wall quantities of interest, i.e., wall heat flux and skin friction. The ensemble Kalman method employed for this purpose provides an effective tool to train the turbulence model using sparse and potentially noisy observation data of mean flow quantities. In this work, spatially sparse observation data of mean flow velocity and temperature at a specific streamwise location along the flat plate are used for training.

The DNS database was used to train the turbulence model and evaluate its performance on test flow cases. The model, trained with observation data from a single flow case, successfully learned and predicted results for that specific case. However, when applied to the test flow cases, the model produced mixed results, with some predictions showing slight improvements and others performing worse than the baseline solution. In contrast, the model trained with data from two flow cases successfully captured the variability in quantity associated to Reynolds stress under different cold--wall conditions. Furthermore, the jointly trained model demonstrated better generalization to the test flow cases, particularly in predicting wall heat transfer and skin friction coefficients, compared to the single-case training model.

The results presented in this work highlight several limitations and key insights that could inform future research. A major limitation is the lack of variation in the turbulent Prandtl number along the wall-normal direction predicted by the NN-based turbulence models, in contrast to the nonlinear variation observed in the DNS data. The turbulent Prandtl number profile from DNS data exhibits significant changes across the inflection point of the temperature profile. Whether this deficiency stems from the training framework or model--form inadequacy remains an open question for future investigation.

Finally, adopting a nonlinear representation of Reynolds stress could offer potential improvements over the current linear approach. Future work will aim to address these limitations and explore various strategies to enhance the generalizability of the learned turbulence model.

\section*{Appendix A. Higher wall--temperature ratio case}
\label{sec:appendixA}
M5Tw091 provides a challenging test case for an NN-based turbulence model trained for cold-wall conditions. Specifically, with a $T_w/T_r$ of 0.91 and a Mach number of 4.9, it tests the extrapolating capability of the jointly trained turbulence model for M6Tw076 and M14Tw018 cases. The freestream conditions and wall temperature for the M5Tw091 case are given in Table~\ref{tab:M5_case}.

\begin{figure}[hbt!]
\centering
\subfloat[Velocity profile]{\includegraphics[width=0.35\textwidth]{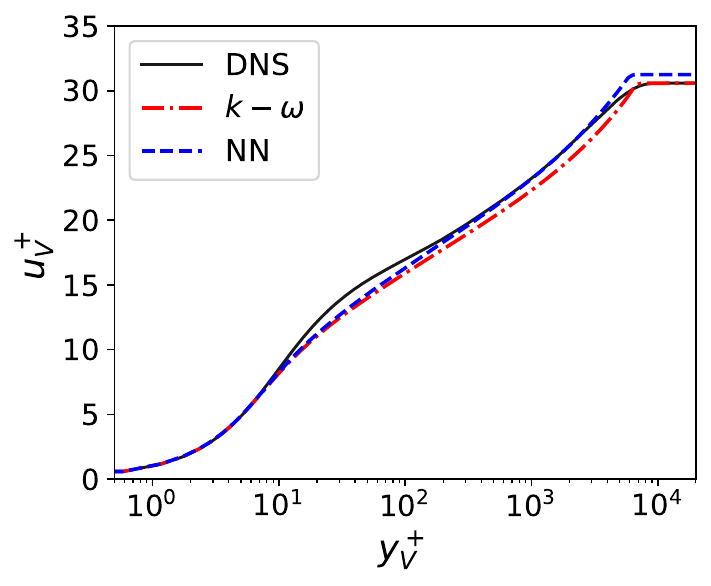}} \
\subfloat[Temperature profile]{\includegraphics[width=0.34\textwidth]{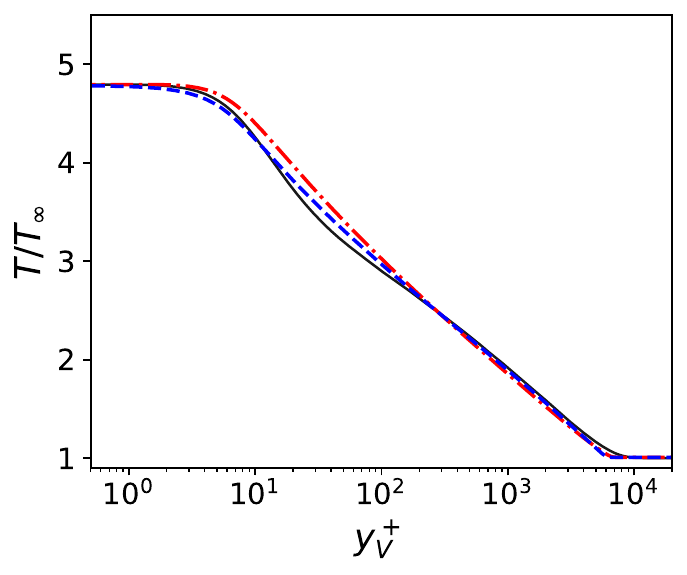}} \
\subfloat[Wall heat transfer Coefficient]{\includegraphics[width=0.35\textwidth]{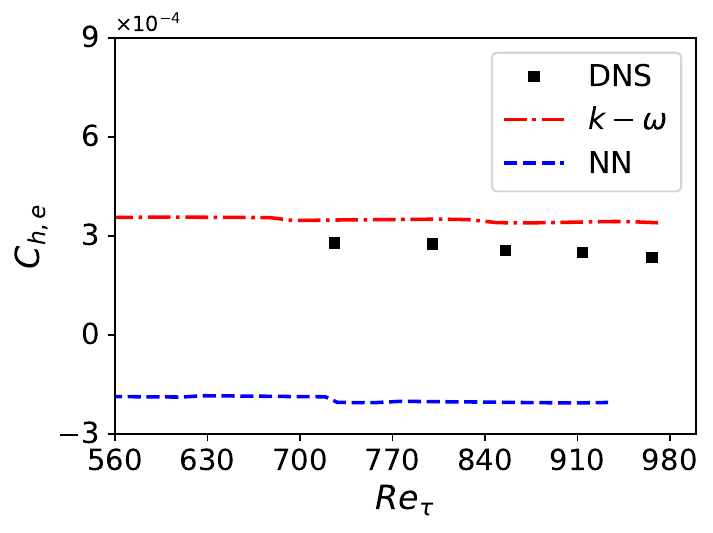}} \
\subfloat[Skin friction coefficient]{\includegraphics[width=0.35\textwidth]{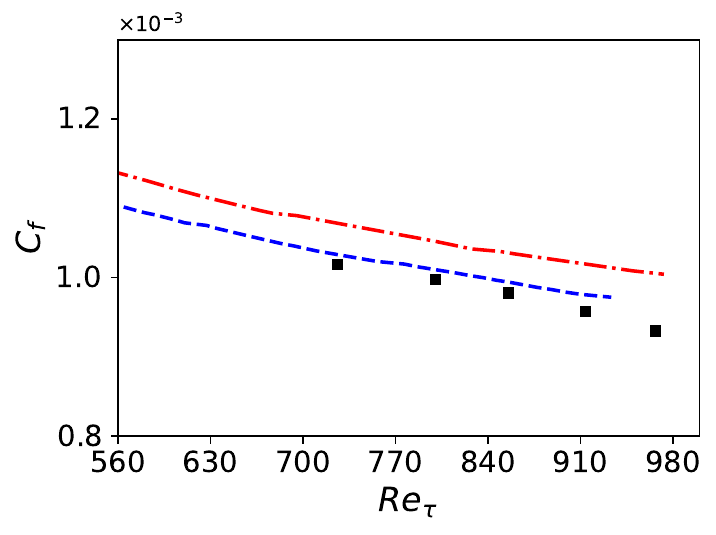}}
\caption{Predicted results for M5Tw091 case, with NN model jointly trained with M6Tw076 and M14Tw018 data}
\label{fig:joint_test_M5Tw091}
\end{figure}

The results are presented in Fig.\ref{fig:joint_test_M5Tw091}. Unlike the cold--wall cases listed in Table\ref{tab:flow_cases}, the heat flux toward the wall in this case is small, and the temperature profile is nearly flat close to the wall. Similar to the observations made for the M6Tw076 case, the baseline turbulence model ($k$--$\omega$) outperforms the trained NN turbulence model in computing the velocity profile toward the boundary layer edge. This re-emphasizes the point that the standard value of $g_1=-0.09$ performs relatively better for $T_w/T_r$ closer to 1, whereas the jointly trained NN turbulence model predicts $g_1$ closer to -0.06 for this test case. Furthermore, the NN turbulence model incorrectly predicts the edge heat transfer coefficient ($C_{h,e}$) as negative, effectively representing heat transfer away from the wall. This aligns with the predicted temperature profile in Fig.~\ref{fig:joint_test_M5Tw091}(b), where the temperature profile near the wall predicted using the NN turbulence model has a slight negative slope, unlike those of the $k$--$\omega$ model and DNS.
\begin{table}[htbp]
\centering
\caption{Freestream conditions and wall temperature for higher wall temperature ratio case \label{tab:M5_case}}
\begin{tabular}{l r r r r r r r c} \hline
Case & $M_\infty$ & $U_\infty, \text{m/s}$ & $\rho_\infty, \text{kg/m}^3$ & $T_\infty, \text{K}$ & $T_w, \text{K}$ & $T_w/T_r$ & $\delta_i, \text{mm}$ & $(x_a-x_i)/\delta_i$ \\
\hline \hline
M5Tw091 & 4.9 & 794.0 & 0.272 & 66.2 & 317.0 & 0.91 & 4.0 & 54.0  \\ \hline
\end{tabular}
\end{table}

\section*{Appendix B. Reynolds Analogy Factor for DNS data}
\label{sec:appendixB}
\begin{figure}[hbt!]
\centering
\includegraphics[width=0.5\textwidth]{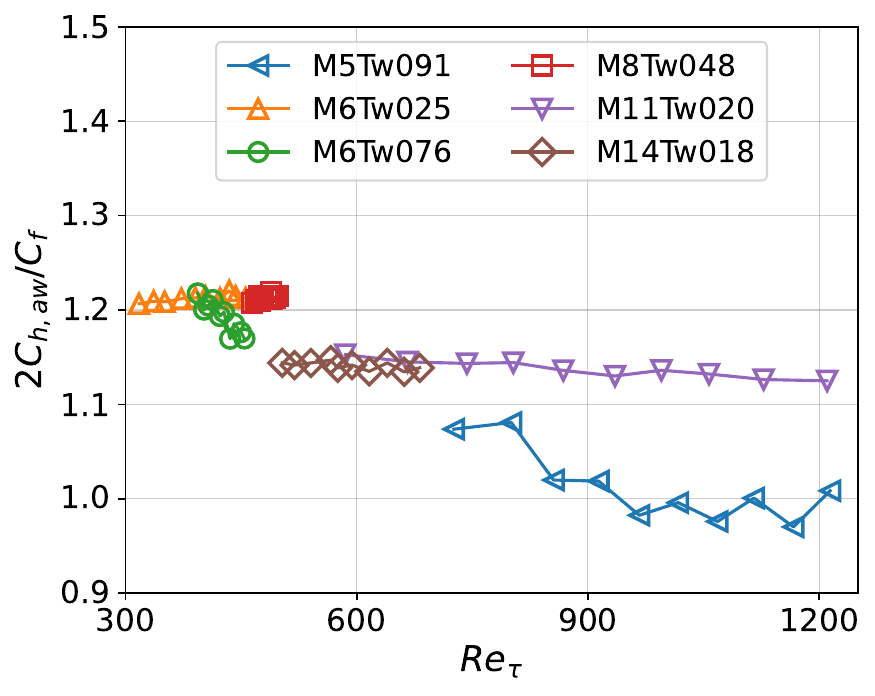}
\caption{Reynolds Analogy factor $R_{af}$ for the DNS data as a function of friction Reynolds number $Re_{\tau}$.}
\label{fig:RA_plots_DNS}
\end{figure}

With a known skin friction coefficient, the Reynolds analogy is often employed to predict the wall heat transfer coefficient~\cite{roy2006pas}. For compressible flows, the Reynolds analogy factor is defined as $ R_{af} = 2C_{h,aw}/C_f $. For the DNS data used in this work, Figure~\ref{fig:RA_plots_DNS} shows the streamwise variation of $R_{af}$ with respect to the Reynolds number based on friction velocity and wall viscosity ($Re_\tau$). Across the range of $Re_\tau$ values covered by the available DNS data~\cite{huang2022dnscases}, $R_{af}$ remains nearly constant, consistent with typical expectations, for all flow cases except M5Tw091 and M6Tw076. In these two cases, where $T_w/T_r$ is closer to 1, mean wall heat flux values are small, resulting in relatively larger statistical errors compared to the other cases.

\section*{Funding Sources}
This research was supported by the Office of Naval Research under the contract number N68335-22-C00271.

\section*{Acknowledgments}
The authors would like to gratefully acknowledge Lian Duan for engaging in valuable discussions, as well as providing the DNS data. The computational resources used for this project were provided by the Advanced Research Computing (ARC) of Virginia Tech, which is also gratefully acknowledged.

\bibliographystyle{apalike}

\end{document}